\documentclass[12pt]{article}
\usepackage[margin=1in, top=1in]{geometry}  % add also "showframe" options
\newfont{\bbf}{cmbx12 scaled 1435}

\usepackage{setspace}
\usepackage{epsfig, graphics}
\usepackage{amsmath,amssymb}
\usepackage[latin1]{inputenc}
\usepackage{graphicx}
\usepackage{epsfig}
\usepackage{amsfonts}
\usepackage{caption}
\usepackage{subcaption}

\newcommand{\nin}{\noindent}
\newcommand{\be}{\begin{equation}}
\newcommand{\ee}{\end{equation}}

\renewcommand{\theequation}{\thesection.\arabic{equation}}

\newcommand{\ba}{\begin{eqnarray}}
\newcommand{\ea}{\end{eqnarray}}
\newcommand{\ban}{\begin{eqnarray*}}
\newcommand{\ean}{\end{eqnarray*}}

\def \1n{1\hskip -3pt \mbox{N}}
\def \eR{I\hskip -3pt R}

\begin{document}
\setlength{\baselineskip}{.26in}
\thispagestyle{empty}
\renewcommand{\thefootnote}{\fnsymbol{footnote}}
\vspace*{0cm}
\begin{center}

\setlength{\baselineskip}{.32in}
{\bbf Nonlinear Fore(Back)casting and Innovation Filtering for Causal-Noncausal VAR Models}\\

\vspace{0.4in}

\large{Christian Gourieroux}\footnote{University of Toronto, Toulouse School of Economics and CREST, e-mail: {\it Christian.Gourieroux @ensae.fr} },
\large{and Joann Jasiak}\footnote{York University, e-mail: {\it jasiakj@yorku.ca}\\
}

\setlength{\baselineskip}{.26in}
\vspace{0.8in}

This version: \today\\

\medskip

\vspace{0.4in}
\begin{minipage}[t]{12cm}

\small

\begin{center}
Abstract \\
\end{center}
We show that the mixed causal-noncausal Vector Autoregressive (VAR) processes
satisfy the Markov property in both calendar and reverse time.
Based on that property, we introduce closed-form formulas of forward and backward predictive densities for point and interval forecasting and backcasting out-of-sample. The backcasting formula is used for adjusting the forecast interval to obtain a desired coverage level when the tail quantiles are difficult to estimate. A confidence set for the prediction interval is introduced for assessing the uncertainty due to estimation. We also define new
nonlinear past-dependent innovations of mixed causal-noncausal VAR models for impulse response function analysis. Our approach is illustrated by simulations and an application to oil prices and real GDP growth rates.

\bigskip

{\bf Keywords:}  Mixed Causal-Noncausal Process, Bubble, Predictive Density, Backcasting, Nonlinear Innovations,  Generalized Covariance (GCov) Estimator, Oil Price.

\end{minipage}

\end{center}
\renewcommand{\thefootnote}{\arabic{footnote}}

\doublespacing
\newpage

\section{Introduction}

The causal-noncausal (mixed) Vector Autoregressive processes have
nonlinear dynamics with locally explosive patterns such as local trends, spikes and bubbles [Gourieroux and Jasiak (2017), Davis and Song (2020)]. In applied research, these strictly stationary models can replicate the behavior of commodity prices, including oil prices [Lof and Nyberg (2017), Cubadda et al. (2023), Blasques et al. (2025)], cryptocurrency rates [Hencic and Gourieroux (2019), Hall and Jasiak (2024), Cavaliere et al. (2020)], financial indexes, including S\&P500 and NASDAQ [Gourieroux, Zakoian (2017), Freis (2022)], climate risk on El Nino and La Nina [De Truchis, Fries and Thomas (2024)], and green stocks prices [Hecq et al (2024)]. Specifically, a  mixed VAR(1) process ($Y_t$) of dimension $m$ is the strictly stationary solution of $Y_t = \Phi Y_{t-1} + \varepsilon_t$, where the errors $\varepsilon_t$ are assumed to be non-Gaussian, independent and identically distributed (i.i.d.), and the eigenvalues of of autoregressive matrix $\Phi$ are of modulus either smaller or greater than 1, and are referred to as the causal and non-causal eigenvalues, respectively. In the presence of non-causal eigenvalues, the errors $\varepsilon_t$ are correlated with the lagged values of $Y_t$ and cannot be interpreted as the (causal) innovations of the process. We show that the mixed VAR model has a causal, i.e. past-dependent representation, which is a multivariate non-linear  autoregression $Y_t = a(Y_{t-1}; v_t)$, where $a$ is a non-linear function, and the errors $v_t$ are i.i.d. and independent of $Y_{t-1}$, satisfying the definition of {\it causal} innovations. 

The initial papers on mixed VARs were focused on the identification of noncausal dynamics and parameter estimation. Davis and Song (2020) introduced the maximum likelihood estimator under a parametric assumption on the distribution of error $\epsilon_t$. Gourieroux and Jasiak (2017), (2023) proposed a semi-parametric one-step Generalized Covariance (GCov) estimator which is consistent and semi-parametrically efficient. 
This paper is focused on the Markov property of mixed VAR processes, which leads to nonlinear predictive density formulas and nonlinear impulse response functions (IRF) for these processes.
We derive the closed-form expressions of forward and backward predictive densities, and build prediction intervals (sets) that account for estimation risk and are adjustable to the desired conditional coverage. We also discuss the identification of the aforementioned causal innovations and define the nonlinear IRF for tracing out the effects of shocks to the locally explosive component of the process.  
The aim of this paper is to provide a toolbox of inference methods for mixed VAR models, equivalent to those available for the traditional causal VARs, to facilitate the use of  these models in applied research in finance and macroeconomics.
 
The closed-form predictive density for forecasting the mixed VAR processes is new and derived in a semi-parametric framework. This closed-form representation was deemed infeasible until recently\footnote{"The predictive density is generally not available under closed form", Fries and Zakoian (2019)}. It extends to a general framework earlier results on a mixed VAR model with a multiplicative form of autoregressive matrix, also called the multivariate Mixed Autoregressive (MAR) model [see Nyberg and Saikkonen (2014), Gourieroux and Jasiak (2016), Lanne and Luoto (2016)]. 
The two types of models are not equivalent, except for pure causal and noncausal processes, because the multiplicative form assumption
is strongly constraining and may not exist for all mixed VAR processes. In particular, it does not exist for a bivariate, mixed VAR(1) model with at least one causal and one non-causal eigenvalues [see Appendix A.2]. The availability of a closed-form predictive density for the general VAR model considered in this paper eliminates 
the need for assuming a restrictive multiplicative form of autoregressive matrix and 
using computationally intense forecasting methods based on numerical approximations. %truncation biases, 

From the quantiles of the predictive densities, we infer the point forecasts and prediction intervals. In addition, we derive a new backcasting method for noncausal processes. The backcasting algorithms are readily available for time-reversible linear Gaussian time series, but need to be developed specifically for the time-irreversible non-Gaussian mixed VAR processes for the treatment of missing data, or  implementation of backpropagation algorithms in applied research
%because the assumption of Gaussianity underlies the commonly used time series and machine learning methods of backcasting 
[Twumasi and Twumasi (2022)]. In our paper, we use the backcasting algorithm to build a new bootstrapped prediction interval adjustable to a desired coverage level, and we show how to account for the estimation risk by introducing the confidence set for prediction intervals.
%The novelty of this approach is in considering the future value of the process as an unknown parameter leading to a partial identification framework. 
This approach is applicable to any nonlinear dynamic model where the forecast error is not a sum of the unknown future innovation and estimation error, and the process is back-castable. The  adjustment of the empirical coverage of a prediction interval to the desired one is beneficial when the multivariate predictive density is estimated from data with a limited number of tail observations, or evaluated over insufficiently support points because of a computational burden. 
%Then, its advantage is in providing prediction intervals at higher levels despite the scarcity of extreme (tail) observations  in finite sample by extending prediction intervals obtained at lower levels.

Another contribution of this paper is
in defining the nonlinear causal innovations for mixed VAR models and examining their identification issues. We describe the nonlinear IRF analysis of mixed VARs, with a shock applied to a locally explosive component of the process, and performed on- and off-bubble. 
 The nonlinear causal innovation for the mixed VAR model is based on
Gourieroux and Jasiak (2005), which has recently received attention in the context of nonlinear impulse response functions in macroeconomics [see, e.g. Gonzalves et al.(2021), Gourieroux and Lee (2025)]. 

The paper is organized as follows. Section 2 describes the causal-noncausal VAR model and its state-space representation. Sections 3, 4 and 5 contain the new results. Section 3 proves the Markov property and derives the closed-form formulas of multivariate predictive density for forecasting and backcasting. Section 4 introduces the inference on the random set of prediction intervals.
Section 5 defines the  nonlinear causal innovations and discusses their identification and filtering. A simulation study and an empirical application to a bivariate process of oil prices and real US GDP rates are presented in Section 6.
Section 7 concludes. The technical results are given in Appendices A.1-A.2. Appendix A.1 contains the proof of the forward and backward predictive density formula. Appendix A.2  explains the constraints induced by the multiplicative representation of a causal-noncausal VAR model. Online Appendices B, C, D and E present
the closed-form expression of a kernel-based semi-nonparametric estimator of predictive density, give additional results on simulations and the empirical application of the mixed VAR model to oil prices and real GDP rates.

\setcounter{equation}{0}\def\theequation{2.\arabic{equation}}

\section{Mixed Causal-Noncausal Processes}
\setcounter{equation}{0}\def\theequation{2.\arabic{equation}}

This Section reviews the state space representation of the causal-noncausal (mixed) VAR(p) model studied in Gourieroux and Jasiak (2016),(2017), and Davis and Song (2020).

\subsection{The Model}
\nin The multivariate causal-noncausal VAR(p), referred to as the mixed VAR process henceforth, is defined by:
\begin{equation}
Y_t  =  \Phi_1 Y_{t-1} + \cdots + \Phi_p Y_{t-p} + \varepsilon_t,
\end{equation}
\noindent where $Y_t$ is a vector of size $m$, $\Phi_j, \, j=1,...,p$, are matrices of autoregressive coefficients of dimension $m \times m$ and $(\varepsilon_t)$ is a sequence of errors, which are  serially i.i.d. random vectors of dimension $m$ with mean zero, %finite variance-covariance matrix $\Sigma$, 
and common joint density $g$. The errors $(\varepsilon_t)$ are assumed to have a non-Gaussian distribution and are not assumed independent of past $Y$'s. This implies that  $(\varepsilon_t)$  cannot be interpreted as the (causal) innovations.

For the existence of a unique strictly stationary solution to the VAR model (2.1),
we assume that the roots of the characteristic equation of the autoregressive polynomial matrix $det(Id - \Phi_1 \lambda - \cdots  \Phi_p \lambda^p) = 0$
are of modulus either strictly greater, or strictly smaller than one, i.e. are either outside, or inside the unit circle, and that the joint density $g$ has a uniform tail index $\alpha$. 
The strictly stationary solution $(Y_t)$  to model (2.1) can be written as an infinite two-sided moving average (MA($\infty$)) in errors $\varepsilon_t$:
\begin{equation}
Y_t = \sum_{j = -\infty}^{+ \infty} C_j \varepsilon_{t-j}.
\end{equation}
\nin This is a linear time series, according to the terminology of Rosenblatt (2012).
The autoregressive matrices $\Phi_1,...,\Phi_p$ and the matrices of coefficients $C_j$ on the past and future terms of this MA representation are uniquely defined when  $\varepsilon_t$ is non-Gaussian, which is an identifying assumption. Process
$Y_t$ is said to be causal in $\varepsilon_t$, if $Y_t = \sum_{j = 0}^{+ \infty} C_j \varepsilon_{t-j}$, noncausal in $\varepsilon_t$,
if $Y_t = \sum_{j = - \infty}^{-1} C_j \varepsilon_{t-j} =  \sum_{j = 1}^{+ \infty} C_{-j} \varepsilon_{t+j}$,
or mixed, otherwise.

In the presence of a noncausal component, the assumption of strict stationarity of ($Y_t$) implies that, in calendar time, $(Y_t)$ has a nonlinear dynamic and past-dependent conditional heteroscedasticity. Then, the process $(Y_t)$ can be characterized by a rather complicated conditional distribution of $Y_{t+h}$ given $\underline{Y_t} = (Y_t, Y_{t-1},...)$ for $h=1,2,..$,
that leads us to the nonlinear out-of-sample (oos) predictive distribution described in Section 3. 
The nonlinearity in the path of $(Y_t)$ manifests itself through local trends, spikes and bubbles similar to those observed in the time series of commodity (oil) prices, exchange rates, or cryptocurrency prices [Gourieroux and Zakoian (2017), Gourieroux and Jasiak (2017), Gourieroux, Jasiak and Tong (2021)].

\subsection{State-Space Representation}

Let us recall the state-space representation of the mixed VAR process  with the latent causal and noncausal components of the observed process as state variables.

\nin a) {\it The mixed VAR(1) representation of a mixed VAR(p) process}

\nin As it is commonly done in the literature on multivariate causal autoregressive processes, model (2.1) can be rewritten as a $n=mp$ multivariate mixed VAR(1) model, by stacking the present and lagged values of process $(Y_t)$:
\begin{equation}
\left(\begin{array}{c} Y_{t} \\ \tilde{Y}_{t-1} \end{array}\right) \equiv \left(\begin{array}{c} Y_{t} \\ Y_{t-1} \\ \vdots \\Y_{t-p+1} \end{array}\right)  = \Psi  \left(\begin{array}{c} Y_{t-1} \\ \tilde{Y}_{t-2} \end{array}\right) +
\left(\begin{array}{c} \varepsilon_{t} \\ 0 \end{array}\right), \;\;
\mbox{with} \;\;
\Psi = \left(\begin{array}{cccc} \Phi_1 & \cdots &  \cdots & \Phi_p \\ Id & 0 &  \cdots &    \\ \vdots & \ddots &
\ddots &  \vdots \\ 0 & \cdots & Id & 0 \end{array}\right). \nonumber
\end{equation}
\nin The eigenvalues of the autoregressive matrix $\Psi$ are reciprocals of the roots of the characteristic equation (2.4).

\nin b) {\it The Change of Basis}

\nin Matrix $\Psi$ has a real Jordan representation:
$
\Psi = A \left( \begin{array}{cc} J_1 & 0 \\ 0 & J_2 \end{array}\right) A^{-1},
$ where $J_1$ (resp. $J_2$) are real $(n_1 \times n_1)$ (resp. $(n_2 \times n_2)$) matrices where $n_2 = n-n_1$,
with
all eigenvalues of modulus strictly less than 1 [resp. strictly larger than 1], and $A$ is a ($n \times n$) invertible matrix [see, Perko (2001), Gourieroux and Jasiak (2017), Section 5.2, for real Jordan representations]. Then, the above equation can be rewritten after the change of basis $A^{-1}$ as:
\begin{equation}
A^{-1} \left(\begin{array}{c} Y_{t} \\ \tilde{Y}_{t-1} \end{array}\right)  = \left( \begin{array}{cc} J_1 & 0 \\ 0 & J_2 \end{array}\right) A^{-1} \left(\begin{array}{c} Y_{t-1} \\ \tilde{Y}_{t-2} \end{array}\right) + A^{-1}
\left(\begin{array}{c} \varepsilon_{t} \\ 0 \end{array}\right).
\end{equation}
\nin Let us introduce a block decomposition of $A^{-1} $:
$
A^{-1} \equiv \left(\begin{array}{c} A^1 \\A^2\end{array}\right),
$ where $A^{1}$ is of dimension $(n_1 \times n)$, and define the transformed variables:
\begin{equation}
Z_t = \left(\begin{array}{c} Z_{1,t} \\ Z_{2,t} \end{array}\right) = A^{-1}  \left(\begin{array}{c} Y_{t} \\ \tilde{Y}_{t-1} \end{array}\right), \;\;\;\eta_t =
\left(\begin{array}{c} \eta_{1,t} \\ \eta_{2,t} \end{array}\right) = A^{-1} \left(\begin{array}{c} \varepsilon_{t} \\ 0 \end{array}\right).
\end{equation}
\nin This leads us to two sets of state components of process ($Y_t$) such that:
\begin{eqnarray}
Z_{1,t} & = & J_1 Z_{1,t-1} + \eta_{1,t}, \nonumber \\
Z_{2,t} & = & J_2 Z_{2,t-1} + \eta_{2,t}.
\end{eqnarray}
\nin Hence, the state-space representation of process ($Y_t$) consists of:
{\bf state equations} (2.5) representing the causal and non-causal dynamics of state variables $Z_t$ and {\bf measurement equations} for $Y_t$ obtained by solving:
$$\left( \begin{array}{c} Y_t \\ \tilde{Y}_{t-1} \end{array} \right) = A Z_t.$$
\nin These measurement equations are deterministic. Therefore, the filtrations generated by processes $(Y_t)$ and $(Z_t)$ are identical.

Since the Jordan representation of matrix $\Psi$ is not unique, the state-space representation is not unique either. It depends on the choice of state variables, i.e. the latent factors $Z_1$ and $Z_2$, up to linear invertible transformations.

\nin c) {\it State-Specific Linear Errors}

\nin The first set of state equations in (2.5) defines the causal VAR(1) process ($Z_{1,t}$), which has the causal MA$(\infty)$
representation:
\begin{equation}
Z_{1,t} = \sum_{j = 0}^{+ \infty} J_1^j \eta_{1,t-j},
\end{equation}
\nin where $\eta_{1,t}$ is a function of $\varepsilon_{t}, \varepsilon_{t-1}...$.
The second set of state equations in (2.5) needs to be inverted to obtain a MA representation in matrices with eigenvalues of  modulus strictly less than 1. We get:
\begin{equation}
Z_{2,t}  =  J_2^{-1} Z_{2,t+1} - J_2^{-1} \, \eta_{2,t+1} 
   =  - \sum_{j = 0}^{+ \infty} [ J_2^{-j-1} \, \eta_{2,t+j+1}].
\end{equation}
\nin We observe that $(Z_{2,t})$ is a noncausal process with a one-sided moving average representation in future values $\varepsilon_{t+1}, \varepsilon_{t+2},...,$. 

Let us now discuss the state-specific linear errors $\eta_{1,t}, \eta_{2,t}$. From equations (2.8), (2.9) and the stationarity conditions, it follows that:
\begin{equation}
\eta_{1,t} = Z_{1,t} - E(Z_{1,t}|\underline{Z}_{1,t-1}), \;\;\; \eta_{2,t} = Z_{2,t} - E(Z_{2,t}|\bar{Z}_{2,t+1}),
\end{equation}
\nin where $\underline{Z}_{1,t-1} = (Z_{1,t-1}, Z_{1,t-2},...$) and $\bar{Z}_{2,t+1} = (Z_{2,t+1}, Z_{2,t+2},...$). Therefore, $\eta_{1,t}$ (resp. $\eta_{2,t}$) can be interpreted as the causal linear innovation of $Z_{1,t}$ based on the information
$\underline{Z}_{1,t-1}$ (resp. noncausal linear innovation of $Z_{2,t}$ based on the information
$\bar{Z}_{2,t+1}$). 
It is important to note that the information set $\underline{Z}_{1,t-1}$ (resp. $\bar{Z}_{2,t+1}$)  differs in general from the global causal information set $\underline{Y}_{t-1} = \underline{Z}_{t-1}$ (resp $\bar{Y}_{t+1} = \bar{Z}_{t+1}$). In Section 5, we clarify this point by introducing the notion of a nonlinear innovation that takes into account all available information.
Before doing that, we need to derive the expressions of forward and backward predictive densities.
\setcounter{equation}{0}\def\theequation{3.\arabic{equation}}

\section{Out-of-Sample Predictive Density}

This Section presents the forecasting and backcasting methods based on closed-form expressions of forward and backward predictive densities. 

\subsection{Forward Predictive Density}

The expression of the predictive density of $Y_{T+1}$ given $\underline{Y}_T = (Y_T, Y_{T-1},...)$ is given below and derived in Online Appendix A.1., assuming that the  matrices of autoregressive coefficients $\Phi_1, \Phi_2,...$ and joint error density $g$ are known.

\nin {\bf Proposition 1:} The conditional probability density function (pdf) of $Y_{T+1}$ given $\underline{Y}_T $ is:
\begin{equation}
l(y| \underline{Y}_T ) = \frac{l_2 \left[A^2 \left(\begin{array}{c} y \\ \tilde{Y}_{T} \end{array}\right)\right]}{l_2 \left[ A^2
\left(\begin{array}{c} Y_{T} \\ \tilde{Y}_{T-1} \end{array}\right) \right]} \, |\det J_2 | \, g(y- \Phi_1 Y_T - \cdots - \Phi_p Y_{T-p+1}),
\end{equation}
\nin where $l_2(z_2)$ is the stationary pdf of $Z_{2,t} = A^2 \left(\begin{array}{c} Y_t \\ \tilde{Y}_{t-1} \end{array}\right)$, if $n_2 \geq 1$. In the pure noncausal process, $n_2=0$, we have:
$ l(y| \underline{Y}_T ) = g(y- \Phi_1 Y_T - \cdots - \Phi_p Y_{T-p+1}).$

In the special case of the mixed VAR(1) process with $p=1$, the predictive density becomes:
\begin{equation} l(y| \underline{Y}_T ) = l(y | Y_T ) = \frac{l_2 (A^2 y)}{ l_2 (A^2 Y_T)} \, |\det J_2|  \, g(y- \Phi Y_T).
\end{equation}

\nin Proof: See Appendix A.1.

\medskip

This predictive density is defined for the VAR representation (2.1). It is a semi-parametric function of the parameters $\Phi_1,...,\Phi_p$, determining $A^2$ and $J_2$, and of the functional parameters $g, l_2$. This predictive density is  complicated, except if it is multivariate Gaussian. In general, the conditional mean is not linear in $\underline{Y}_{T}$. 

\nin From equation (3.1), the deterministic relation (2.4) between $Y_t$ and $Z_t$,
and the symmetry of  the calendar and reverse time scales, it follows that:

\medskip

\nin {\bf Corollary 1: Markov Property} The mixed VAR(p) process  $(Y_t)$ of dimension $m$ [resp. the state process $(Z_t)$
of dimension $n=mp$] is a Markov processes of order $p$ [resp. of order 1] in  calendar time for $n_2 \geq 1$. The processes $(Y_t)$ and $(Z_t)$ are both Markov of orders $p$ and 1, respectively, in reverse time too.

\medskip

\nin This corollary extends to 
%linear mixed, i.e. causal-noncausal processes of 
any autoregressive order $p$ 
the result of Cambanis and Fakhre-Zakeri (1994), who show that a linear pure noncausal autoregressive process of order 1 is a causal Markov process of order 1. 
It also extends the Proposition 3.1 in Freis and Zakoian (2019) derived in the univariate case.

The predictive density summarizes the nonlinear causal dynamics of a mixed VAR process.
From the predictive densities, we infer the point forecasts and prediction intervals based on its quantiles. These lead us to the oos point predictions and prediction intervals at various horizons for this nonlinear and non-Gaussian process analogous to the linear pointwise predictions and prediction intervals of the causal VAR models.

\subsection{Backward Predictive Density}

Since the mixed VAR process of order $p$ is Markov of order $p$ both in calendar and reverse times, we can derive from Proposition 1 the closed-form expression of backward predictive density in reverse time for backcasting. For ease of exposition, we present it below for $p=1$.

\medskip

\nin {\bf Corollary 2: Backcasting} Let us consider a mixed VAR(1) model. The backward predictive density of $Y_{T-1}$ given $Y_T$ is:
$$l_B (y|Y_T) = \frac{l_1(A^1 y)}{ l_1(A^1 y_T)} |det \; J_2| \; g(Y_T - \Phi y),$$
\nin where $l_1$ is the stationary density of $Z_{1t}$ and $g$ is the density of $\varepsilon$.

\nin Proof: See Online Appendix A.1. 

By considering jointly Proposition 1 and Corollary 2, we see that the mixed VAR models have a nonlinear dynamic structure extending to nonlinear dynamic framework the standard Kalman filter available for linear Gaussian processes.

\medskip
In Section 2.3, we mentioned that there exists a multiplicity of real Jordan representations of matrix $\Psi$. It implies a multiplicity of matrices $A$ built from its extended real eigenspaces (in the presence of complex conjugate eigenvalues). However,
$\det J_2 = \prod_{j=1}^{n_2} \lambda_j$, where $|\lambda_j| > 1, \, j=1,...,n_2$, is independent of the real Jordan representation.
Similarly, the noncausal component $Z_2$ is defined up to a linear invertible transformation. Since the Jacobian  is the same for the numerator and denominator of the ratio
$l_2 \left[A^2 \left(\begin{array}{c} y \\ \tilde{Y}_{T} \end{array}\right)\right] \; /\;l_2 \left[ A^2
\left(\begin{array}{c} Y_{T} \\ \tilde{Y}_{T-1} \end{array}\right) \right]$,
it has no effect on the ratio. Thus, the expression of $l(y|\underline{Y}_T)$ does not depend on the selected real Jordan representation, that is on the selected state-space representation. The same remark applies to the backward predictive density.

\subsection{Prediction at Horizon $h$}

 The closed-form expressions of the backward and forward  predictive densities at horizon 1 can be used sequentially to forecast or back-cast out-of-sample at any horizon $h>1$. Alternatively, the predictive densities at horizons $h>1$  can be obtained by using the Sampling Importance Resampling (SIR) method. Then, the forward predictive density at horizon $h$ can be written as a multivariate integral over $h$ future values of the process  and approximated for given $\Phi_1,..,\Phi_p$ and $g$ by drawing  the future values of the process from the closed-form predictive density at horizon 1 replicated $S$ times
by the SIR method [Gelfand and Smith (1992), Tanner (1993)] (see Section 6 for an illustration). Specifically, to approximate the predictive density at horizon $h>1$, for given $\Phi_1, \Phi_2,...$ and $g$,  we need S independent future paths of the process. Each path $s=1,...,S$ is obtained by forecasting sequentially $Y_{T+1}^s| Y_T$, followed by $Y_{T+2}^s| Y_{T+1}^s,...,Y_{T+h}^s|Y_{T+h-1}^s$, from the predictive densities. This becomes a drawing $Y_{T+1}^s,...,Y_{T+h}^s$ of a future path $s$. By replicating it independently for $s=1,...,S$, we get 
$Y_{T+h}^s, \; s=1,...,S$. Then, for a large $S$, we can use the sample distribution 
of  $Y_{T+h}^s, \; s=1,...,S$ as an estimator of the predictive distribution at horizon $h$.

%The closed form of predictive density (3.1) extends to a general framework the expression derived in Gourieroux and Jasiak (2016) for univariate mixed processes, or for multiplicative mixed multivariate models of Lanne and Saikkonen (2011), (2013). Online Appendix A.2. describes the limitations of the multiplicative form of mixed VAR models.  

\setcounter{equation}{0}\def\theequation{4.\arabic{equation}}

\section{Statistical Inference}

The parameters of a mixed VAR model need to be estimated before the forecasts are computed.
Below, we review the existing estimation methods in the time domain and describe the predictive algorithm providing the estimated predictive densities in a semi-parametric setup. 
As pointed out in Section 3, from the quantiles of the estimated predictive density we obtain the point and interval forecasts, which can be impacted by the preliminary estimation step involving estimators converging at different rates. Therefore, in the second part of this section we study the forecast interval uncertainty, using 
the theory of random sets [see Molchanov and Molinari (2018)]. 

\subsection{Estimation and Filtering}

The mixed VAR model can be estimated by the maximum likelihood method based on an assumed parametric distribution of $\varepsilon_t$ [see Breidt et al. (1991), Lanne and Saikkonen (2013), Davis and Song (2020)]. This approach yields consistent estimators provided that the parametric  distributional assumption is correct (and non-Gaussian for identification).

Alternatively, the mixed VAR model can be consistently estimated without any parametric assumptions on the error distribution \footnote{Except for the non-Gaussianity assumption and the uniform tail parameter.} by using the semi-parametric (Generalized) Covariance  (GCov) estimator [Gourieroux and Jasiak (2023)].
The GCov estimator is consistent, asymptotically normally distributed and semi-parametrically efficient. As an alternative in the frequency domain, minimum distance estimators based on the cumulant spectral density of order 3 and 4 have been proposed in Velasco and Lobato (2018) and  Velasco (2022).

\medskip

The prediction  methods introduced in Section 3 for given $\Phi_1,...,\Phi_p$ and $g$  can be applied in a parametric or semi-parametric framework by replacing these unknown parameters by their consistent estimates. In the semi-parametric framework, this can be done along the following lines:

\nin step 1. Apply the GCov estimator based on zero auto-covariance conditions of nonlinear error functions to obtain the estimators of matrices of autoregressive coefficients
$\hat{\Phi}_1, ..., \hat{\Phi}_p$.

\nin step 2. Use the $\hat{\Phi}_i, i=1,...,p,$ estimates to compute the roots of the lag-polynomial and more generally an estimated real Jordan representation: $\hat{A}, \hat{J}_1, \hat{J}_2$.

\nin step 3. Compute the approximated model errors using the estimates obtained in Step 1:\linebreak
$\hat{\varepsilon}_t = Y_t - \hat{\Phi}_1 Y_{t-1} -... - \hat{\Phi}_p Y_{t-p}$.

\nin step 4. Compute $\hat{Z}_t = \hat{A}^{-1} \left(\begin{array}{c} Y_t \\ \tilde{Y}_{t-1} \end{array}\right), \;\;\;
\hat{\eta}_t = \hat{A}^{-1} \left(\begin{array}{c} \hat{\varepsilon}_t \\ 0 \end{array}\right)$.

\nin step 5. The following densities can be estimated by kernel estimators applied to the approximated series:

- the density $g$ of $\varepsilon_t$ can be estimated from  $\hat{\varepsilon}_t, \; t=1,...,T$;

 - the density $l_2$ of $Z_{2,t}$ can be estimated from  $\hat{Z}_{2,t}, \; t=1,...,T$ (see Online Appendix B);

\nin step 6. The predictive density can be estimated from the formula (3.1) by replacing $l_2$ by $\hat{l}_2$, $A^2$ by
 $\hat{A}^2$, and also $J_2$ by $\hat{J}_2$, $g$ by $\hat{g}$, and $\Phi_1, ..., \Phi_p$ by $\hat{\Phi}_1, ..., \hat{\Phi}_p$.
The mode (median) of the predictive density provides the point forecasts and the quantiles of the estimated predictive density can be used to obtain estimated prediction intervals at horizon 1 (see, Online Appendix B).

\subsection{Estimated Prediction Interval Uncertainty}

The estimated model parameters and residuals $\hat{\varepsilon}_t$ can be used to build oos predictions and prediction  intervals conditional on given values of the
last observations in the sample, called the conditional prediction interval. In finite sample, there is uncertainty on the estimated prediction intervals resulting from semi-parametric and non-parametric estimators with different convergence rates. The estimation errors of the scalar and functional parameters have a nonlinear effect on this uncertainty, which should not be disregarded.
%in nonlinear dynamic models.
%Hence it is important to take into account the so-called estimation risk for building prediction intervals with adequate (conditional) coverage in nonlinear dynamic models estimated semi-parametrically.

To highlight the specificities of the nonlinear dynamic model analysis, we compare our approach with the standard practice of predicting from a linear AR(1) model defined below:

{\it Example 1: Linear AR(1) model:} The model is given by: $y_t = \phi y_{t-1} + \epsilon_t, \; |\phi|<1$
where the errors $\epsilon_t$ are i.i.d. with the unknown true density $f_0$.

\nin Below, we compare the true and estimated prediction intervals, and introduce the bootstrap-adjusted prediction intervals.

\nin{\bf 4.2.1 True and Estimated Prediction Intervals}

\nin i) {\bf True Prediction Interval:} For ease of exposition,
let us consider the VAR(1) model, a forecast horizon $h=1$ and a future value of the first component series $Y_{1,T+1}$ to be forecast at date $T$ out of sample (oos)  given $Y_T = (y_{1,T}, y_{2,T})' \equiv y$, where $(Y_{2,t})$ contains the remaining components of the series. Then, the true prediction interval at level $1-\alpha_1$ for $Y_{1,T+1}$ is:
\begin{equation}
PI(y, \alpha_1) = [ Q_l (y,\alpha_1; P_{0}),\, Q_u (y, \alpha_1; P_{0})],
\end{equation}
\nin where $P_{0}$ is the true density function of process ($Y_{t}$) and $Q(y, \alpha, P_0)$ denotes the $\alpha$-quantile of $Y_{1,T+1}$ conditional on $Y_T=y$, derived from the joint
multivariate predictive density $l(y | \underline{Y}_T)$ (see Proposition 1 for the closed-form expression of the predictive density). Let $Q_l$ and $Q_u$ denote the true $\alpha_1/2$ and $1-\alpha_1/2$ conditional quantiles, respectively. Recall that under the semi-parametric approach, the true density
$P_{0}$ is characterized by the parameter matrix $\Phi_{1,0}$ and functional parameter $g_0$. 
Then, conditional on the information at time $T$, the prediction interval is $PI(Y_T, \alpha_1)$. This is a random interval (set), a function of $Y_T$, and its conditional coverage level is
$P_0 (Y_{1, T+1} \in PI(Y_T, \alpha_1)|Y_T) = 1- \alpha_1, \forall Y_T. $
%\nin is equal to $1- \alpha_1$ for any value of $Y_T$.

{\it Example 1: Linear AR(1) model, cont.:} The theoretical prediction interval is: $PI(y, \alpha_1) = ( \phi y - Q_0 (\alpha_1/2), \phi y + Q_0 (1-\alpha_1/2)),$ where $Q_0$ is the quantile function corresponding to the true distribution function of $\epsilon_t$.

In the mixed VAR model $\epsilon_t$ is no longer a causal innovation and the marginal quantiles $Q_0$ have to be replaced by the conditional quantiles $Q(y, \alpha_1, P_0)$ depending on the current value $Y_T=y$. Moreover, this conditional quantile is not an affine function of $y$. 

By using the standard expression of a Gaussian prediction interval with $\Phi$ denoting the Gaussian cumulative distribution function (c.d.f), the asymptotically valid prediction interval (4.1) for $Y_{1,T+1}$  can be equivalently written as:
\begin{equation}
PI(y, \alpha_1) = [m(y,\alpha_1; P_{0}) \pm \Phi^{-1}(\alpha_1/2) \, \sigma (y,\alpha_1; P_{0})],
\end{equation}
\nin with\footnote{Note that $\Phi^{-1}(\alpha_1/2)$ is negative.}
$
m(y, \alpha_1; P_{0}) = 0.5[Q_l (y,\alpha_1 P_{0})+ Q_u (y,\alpha_1; P_{0})]
$ and $\sigma(y,\alpha_1; P_{0}) = - [1/(2\Phi^{-1}(\alpha_1/2)] $

\noindent $[Q_u (y, \alpha_1; P_{0})- Q_l (y, \alpha_1; P_{0})]$.
This normalized representation of prediction interval resembling the traditional Gaussian approach can be used even if the
conditional density function of $(Y_t)$ is not Gaussian. In particular, the functions $ m(y, \alpha_1; P_{0})$ and $ \sigma(y, \alpha_1; P_{0})$ are nonlinear in $y$, in general.

\medskip
\nin {\bf ii) Estimated Prediction Interval:} The unknown marginal predictive density function $P_0$ can be consistently estimated from eq. (3.1) and denoted by $\hat{P}$, given the estimated matrix of autoregressive parameters $\hat{\Phi}_1$ and the residuals $\hat{\varepsilon}_t$ obtained from the semi-parametric GCov estimator, along with the nonparametric estimator $\hat{g}$.  Then, the estimated prediction interval for $Y_{1,T+1}$ is:
\begin{equation}
\widehat{PI}(y, \alpha_1) = [m(y, \alpha_1; \hat{P}) \pm \Phi^{-1}(\alpha_1/2)  \, \sigma (y, \alpha_1;\hat{P})] = [Q_l (y, \alpha_1; \hat{P}),
Q_u (y, \alpha_1; \hat{P})],
\end{equation}
\nin where $Q_l(y, \alpha_1; \hat{P})$ and $Q_u(y, \alpha_1; \hat{P})$ are the $\alpha_1/2$ and $1-\alpha_1/2$ 
conditional quantiles of the estimated predictive density.
This estimated prediction interval (4.3) is consistent of the true prediction interval (4.2) when the number of observations tends to infinity. We know that 
%a) the true prediction interval $PI(Y_T,\alpha_1)$  of $Y_{1,T+1}$ satisfies:
%$P_0 [Y_{1, T+1} \in PI(Y_T,\alpha_1) | Y_T] = 1-\alpha_1, \;\; \forall Y_T.$
%By construction, the true prediction interval has the correct conditional coverage probability of $1-\alpha_1$. It depends on the unknown 
%true distribution, or equivalently on the parameters $\Phi_{1,0}, g_0$;
the estimated prediction interval $\widehat{PI}(Y_T,\alpha_1)$ does not always satisfy the conditional coverage condition in finite sample:
$P_0 [Y_{1, T+1} \in \widehat{PI}(Y_T, \alpha_1) | Y_T] = 1-\alpha_1, \;\; \forall Y_T$. In general, we have:
$P_0(Y_{1, T+1} \in \widehat{PI}(Y_T, \alpha_1)|Y_T) = 1 - \alpha_1 (Y_T),$
i.e. the conditional coverage depends on $Y_T$, and the unconditional coverage:
$P_0(Y_{1, T+1} \in \widehat{PI}(Y_T, \alpha_1)) = E_0[1 - \alpha_1 (Y_T)],$
differs from $1-\alpha_1$\footnote{Because the process is Markov of order 1, we condition on $Y_T$ only instead of $\underline{Y}_T = (Y_T, Y_{t-1},...)$, even though the estimators depend on the entire past.}. 
%due to the estimation errors.

{\it Example 1: Linear AR(1) model, cont.:} The estimated prediction interval becomes:
$\widehat{PI}(Y_T, \alpha_1) = [\hat{\phi} Y_T - \hat{Q}(\alpha_1/2), \hat{\phi} Y_T + \hat{Q}(1-\alpha_1/2)].$
In the linear dynamic model, the estimated quantiles are easily derived from the residuals $\hat{\epsilon}_t = y_t - \hat{\phi}y_{t-1}$ ranked in increasing order.

Such a simple non-parametric estimation method of the quantile function is not available in the nonlinear dynamic framework, including the mixed VAR(1) model, where $\epsilon_t$ is no longer a causal innovation. Instead, the conditional quantiles are estimated from the kernel functional estimator of the predictive density. Note also that the estimates of $Q$ are expected to be less accurate than the estimate of the autoregressive coefficient $\phi$, because they approximate non-parametrically the tails of a multivariate distribution.

\medskip

\nin iii) {\bf 4.2.2 Backward Bootstrap Adjusted Prediction Interval}

Since the estimated prediction interval is random,  
its finite sample distribution can be approximated by a "backward" bootstrap, i.e. by replicating
the trajectory of the process by backcasting, conditional on $Y_T$ to adjust for the bias in coverage conditional on $Y_T$. More precisely, given $\hat{\Phi}_1$, $\hat{g}$ considered fixed and the residuals, we can generate by backcasting the artificial paths $Y_t^s, \; t=1,...,T$,
 with the same terminal condition $Y_T^s=Y_T=y$ for all the bootstrapped samples (see, Corollary 2 for the closed-form expression of the backward predictive distribution). We propose the following algorithm:
 
\nin step 1:  Starting from $Y_T$, we backcast $Y_{T-1}^s$, conditional on $Y_T$, next we backcast $Y_{T-2}^s$ conditional on $Y_{T-1}^s$, and so on. 

\nin step 2: By replicating the backcasted path $S$ times, we end up generating $S$
bootstrapped series $Y_t^s, t=1,...,T$, of length $T$ equal to the length of the initial series and with the same terminal value $Y_T$. 

\nin step 3: From each replicated path $(Y_t^s, t=1,...,T)$,
we estimate the model parameters $\Phi^s$ and $\hat{g}^s$, $s=1,...,S$. This allows us for computing at $Y_{T}$ the predictive density estimator $\hat{P}^s, s=1,...,S$ of $Y_{T+1}$ given $Y_T$ and $S$ new prediction intervals for $Y_{1,T+1}$ from each of the replicated paths.

\nin step 4: The bootstrapped prediction interval obtained from a replicated path is:
\begin{equation}
\widehat{PI}^s(y,\alpha_1) = [m(y,\alpha_1; \hat{P}^s) \pm \Phi^{-1}(\alpha_1/2)  \sigma (y,\alpha_1; \hat{P}^s)],
\end{equation}
\nin where $\hat{P}^s$ is the semi-parametric estimate of $P_{0}$ from the generated path $(Y_t^s, t=1,...,T)$.
This bootstrap
PI of $Y_{1,T+1}$ given $Y_T$ can be replicated independently $S$ times.
The components of the prediction interval (4.4) are denoted by:
\begin{equation}
\hat{m}^s (y,\alpha_1) = m(y,\alpha_1; \hat{P}^s), \;\; \hat{\sigma}^s (y,\alpha_1) = \sigma(y,\alpha_1; \hat{P}^s), \; s=1,...,S.
\end{equation}

\nin step 5: For large $S$, the joint sample distribution of $[\hat{m}^s (y,\alpha_1), \hat{\sigma}^s (y,\alpha_1)]$ provides an approximation of the conditional distribution of $[m(y, \hat{P}), \sigma(y,\hat{P})]$ given $Y_T$, when $T$ is finite and sufficiently large.

{\it Example 1: Linear AR(1) model, cont.:} In the linear AR(1) model, the bootstrap is usually applied by drawing independently in the sample distribution of residuals. This is justified by the additive decomposition of the conditional quantile. 

In a nonlinear dynamic model with nonlinear dependence between $y$ and $Q$, the bootstrap has to be performed 
conditional on $Y_T$ to adjust for the bias in coverage conditional on $Y_T$.

\subsection{ Confidence Set of the Prediction Interval}

The estimated prediction interval $\widehat{PI}(y,\alpha_1)$ is a pointwise estimator of interval $PI(y,\alpha_1)$ and, as any estimator, is random itself. This randomness is difficult to assess since $\widehat{PI}(y, \alpha_1)$ is a random interval [see Molchanov and Molinari (2018) for random sets in Econometrics].
Let us now extend the method of pointwise estimation to build confidence sets for $PI(y, \alpha_1)$. Since there does not exist a total ordering on intervals, we constrain the confidence set to be also of the form:
\begin{equation}
\widehat{PI}(y,\alpha_1,q) =  [m(y,\alpha_1; \hat{P}) \pm q \, \sigma (y, \alpha_1;\hat{P})],
\end{equation}
\nin and search for an estimator of $q$ providing the correct asymptotic coverage.  This confidence set has the following conditional coverage probability of the true prediction interval:
\vspace{-0.1in}$$
\begin{array}{lcl}
\Pi_0 (y,\alpha_1; q) & = & P_0 [ \widehat{PI}(y,\alpha_1,q) \supset PI(y,\alpha_1)| Y_T=y] \\
& = & P_0 [m(y,\alpha_1; \hat{P}) - q \, \sigma (y, \alpha_1;\hat{P}) < m(y,\alpha_1; P_0) + \Phi^{-1}(\alpha_1/2)  \, \sigma(y, \alpha_1;P_0),  \\
m(y,\alpha_1; \hat{P}) &  + & q \, \sigma (y,\alpha_1; \hat{P}) > 
m(y,\alpha_1; P_0) - \Phi^{-1}(\alpha_1/2)  \, \sigma(y,\alpha_1; P_0)|Y_T=y] \hfill(4.7)
\end{array}$$

\setcounter{equation}{7}\def\theequation{4.\arabic{equation}}
\nin because $\Phi^{-1}(\alpha_1/2) <0$. For $\alpha_2 \in (0,1)$, possibly different from and less than $\alpha_1$, there exists a value $q_0(y, \alpha_1; \alpha_2)$ such that:
\begin{equation}
\Pi_0[y, \alpha_1;q_0(y, \alpha_1;\alpha_2)] = 1-\alpha_2, \; \forall y.
\label{e2}
\end{equation}
\nin Asymptotically, we get the conditional $1-\alpha_2$ coverage probability  of the true conditional prediction interval, although the true predictive density $P_0$ and the true distribution of $\hat{P}$ remain unknown. In practice, this procedure can provide us a prediction interval at level 95\%, for example, when the available sample does not contain sufficiently many tail observations and a reliable prediction interval at level 90\% or lower is only estimable. 

Then, equations (4.7) and (\ref{e2}) can be replaced by their  bootstrapped counterparts obtained from $S$ replicated paths of the series, which are backcast prior to T, given $Y_T=y$. Next, a forecast at T+1 is computed from each of the replicated paths.
More precisely, the bootstrapped conditional coverage probability is defined as:
$$
\hat{\Pi}^s(y,\alpha_1,q) = \frac{1}{S} \sum_{s=1}^S \delta^s,
$$ 
where
$\delta^s = \left\{ 
\begin{array}{ll}
 1, & \mbox{if} \; \hat{m}^s(y,\alpha_1) - q  \, \hat{\sigma}^s (y,\alpha_1)  <
\hat{m}(y,\alpha_1) + \Phi^{-1}(\alpha_1/2)   \,  \hat{\sigma} (y,\alpha_1), \\
&\mbox{and}\;\hat{m}^s(y,\alpha_1) + q  \,  \hat{\sigma}^s (y,\alpha_1)  >
\hat{m}(y,\alpha_1) - \Phi^{-1}(\alpha_1/2)   \, \hat{\sigma} (y,\alpha_1), \\
0, & \mbox{otherwise}.
\end{array}
 \right.
$

\medskip
\nin Then, we consider a solution $\hat{q}^S(y,\alpha_1, \alpha_2)$ of:
\begin{equation}
\hat{\Pi}^s [y,\alpha_1, \hat{q}^S (y,\alpha_1, \alpha_2)] = 1-\alpha_2.
\end{equation}
\nin ensuring a $1-\alpha_2$ conditional coverage probability. The bootstrap confidence set for the prediction interval is:
\begin{equation}
\widehat{CSPI}(y, \alpha_1,\alpha_2) = \left\{ m(y,\alpha_1, \hat{P}) \pm \hat{q}^S (y,\alpha_1, \alpha_2) \sigma(y,\alpha_1, \hat{P}) \right\},
\end{equation}
\nin and
$
\lim_{T \rightarrow \infty} \lim_{S \rightarrow \infty} P_0 [ \widehat{CSPI}(y, \alpha_1, \alpha_2) \supset
PI (y,\alpha_1)$
$|Y_T=y] = 1-\alpha_2, \; \forall y, \forall P_0.
$
Hence, the length of the estimated $\widehat{PI} (y,\alpha_1)$ is modified by a factor $\hat{q}^S (y,\alpha_1, \alpha_2)/|\Phi^{-1}(\alpha_1/2)| $, that depends on the observed value $Y_T=y$, in general. In practice, we can choose $\alpha_1=\alpha_2 = 0.05$ corresponding to the standard levels for prediction and confidence intervals, respectively. As mentioned earlier, we can choose $\alpha_1$ different from $\alpha_2$, including  $\alpha_1 = 1.0$, which would correspond to the confidence set for a point prediction equal to the median of the predictive density. 
%This method allows for building prediction intervals at high levels by adjusting  prediction intervals at lower levels obtained from data containing insufficient number of tail observations, or when the predictive density is evaluated over a coarse support. 

The analysis of confidence set for the prediction interval is related to set identification, where a confidence set for the identified set is determined for models under partial identification [Beresteanu et al., (2017)]. The partial identification literature considers a parametric model with a partly identifiable parameter\footnote{See Imbens and Manski (2004) for confidence intervals of identified intervals in the framework of partial identification and confidence intervals that asymptotically cover the true interval with a probability larger or equal to $1-\alpha_2$.} to determine either the confidence set under the classical approach, or the credible set under the Bayesian approach. In our framework, we have two types of "parameters": $P$ and $Y_{1, T+1}$. The second one is not identifiable, although its conditional distribution is estimated, and plays the role of a conditional prior.

\setcounter{equation}{0}\def\theequation{5.\arabic{equation}}

\section{Nonlinear Causal Innovations}

The inference on the traditional causal VAR model
commonly includes the IRF analysis, which is an  important tool used by economists and financial policy makers for the so-called causal analysis. This terminology differs from the "causal-noncausal" terminology introduced by Rosenblatt for time series. More precisely, the "causal analysis" involves a) making inference on the future given the past, i.e. forecasting, and also b) the counterfactual analysis of the effects of current transitory shocks on the future. 
The standard linear causal IRF analysis
cannot be applied to the mixed VAR(p) processes, because neither errors $\varepsilon_t$ in model (2.1), nor the state-specific linear errors $\eta_t$ in (2.5) are causal and independent of the lagged values of $Y_t$. Hence, it is difficult to interpret the shock on $\varepsilon_t$ that implicitly changes a past that has already been realized. This section introduces an alternative concept of innovation that eliminates such a difficulty in nonlinear IRF analysis. 
%the presence of noncausal roots in the mixed VAR model implies nonlinear causal dynamics, characterized by bubbles, for example. 
This new innovation accounts for nonlinear dynamics of $(Y_t)$ with bubbles and spikes.
In addition, it satisfies both the serial and cross-sectional independence conditions [Gourieroux and Jasiak (2005), Gourieroux, Monfort and Renne (2017) on structural SVAR models in Macroeconomics].  We discuss below the filtering and  identification of the nonlinear causal innovations in mixed VAR models.

\subsection{Definition of Nonlinear Causal Innovations}

The  nonlinear causal innovations  are defined below for any Markov process of order $p$, including any  mixed VAR(p) model.

\medskip

{\bf Definition 1:} Let us consider a Markov process of order $p$ with a positive continuous transition density. A nonlinear causal innovation of process $(Y_t)$ is a process $(v_t)$ of dimension $m$ such that:
i) the vectors $v_t$ are serially i.i.d.; ii) the strictly stationary process $(Y_t)$ can be written in a nonlinear causal autoregressive form:
\begin{equation}
Y_t = a(\underline{Y_{t-1}}, v_t), \; \forall t,
\end{equation}
\nin with $\underline{Y_{t-1}} = (Y_{t-1},...,Y_{t-p})$;
iii) the future values $\bar{v}_t$ of the innovation process are independent of the  lagged values $\underline{Y_{t-1}}$ of the observed process $(Y_t)$.

It is easy to see that conditions i) and ii) are equivalent to the Markov of order $p$ property of $(Y_t)$ with a continuous distribution, and can be applied in particular to the mixed model (2.1) by Corollary 1. Such a nonlinear causal autoregressive form always exists [see Rosenblatt (1952) and the discussions in Sections 5.2, 5.3]. It is not unique in a nonparametric framework since $v_t$ can always be replaced by $w_t = d(v_t)$, where $d$ is a diffeomorphism, and $a$ is replaced by $a \circ d^{-1}$. If function $a$ is invertible with respect to $v$, then  $v_t$ is a nonlinear function of $Y_t$ given its past and condition iii) is satisfied. Under conditions i) and ii) the autoregressive equation (5.1) can be applied recursively.

%The nonlinear autoregressive model (5.1) is a nonlinear causal representation of the mixed VAR.
The sequence of nonlinear causal innovations provides the basis of nonlinear impulse response functions [Gourieroux and Jasiak (2005), Gonzalves et al. (2021)].
Let us consider a transitory shock $\delta$ at date $T$ on $v_T$, and then apply recursively the autoregressive equation (5.1) to get the IRF, defined as shocked future values $Y_T^{\delta} = a(Y_{T-1}, Y_{T-2},\cdots, v_T + \delta)$, $Y_{T+1}^{\delta} = a( Y_{T}^{\delta}, Y_{T-1},..., v_{T+1})$,  and so on.
The conditions i) and iii) in Definition 1 are crucial for the interpretation of these shocks in terms of "causal analysis". Condition iii) means that $v_T$ can be shocked at any time T without an effect on the realized past $\underline{Y}_{T-1}$.  Condition i) means that this shock has no effect on the future values $\bar{v}_{T+1} = (v_{T+1}, v_{T+2},...)$.

The nonlinear IRF defined above corresponds to a multivariate shock $\delta$. The literature on "causal analysis" is also interested in univariate shocks to specific variables. It is possible to define shocks to a specific component $v_{1,t}$  of $v_t$, if additionally the component $v_{1,t}$ and ($v_{2,t},...,v_{n,t}$) are cross-sectionally independent. 

\subsection{Identification of Nonlinear Autoregression and Innovation}

In the linear causal VAR model, there are identification problems concerning the parameters of the model and the innovation-based IRFs, because the innovations to be shocked are defined up to an orthogonal linear transform. Both these identification problems are solved by identifying the mixing matrix $D$ in the $\varepsilon_t=D v_t$ error representation, using the Independent Component Analysis (ICA) method for example [see  Gourieroux, Monfort and Renne (2017)].
% We get a nonlinear dynamic autoregressive model where the distribution of $v_t$ can be fixed arbitrarily. Then, the parameter identification is solved by assuming a non-Gaussian error distribution, while T
%The identification of a causal innovation remains a problem, discussed below.

The nonlinear autoregression (5.1) depends on function $a$ and the distribution 
of the causal nonlinear innovation $v_t$. In the mixed VAR(p) model, both depend on $\Phi_1, \Phi_2,...$ and $g$, without being one-to-one functions of  these parameters, because the nonlinear function $a$ depends on more arguments than the density $g$. 
This new nonparametric identification issue concerning the causal innovation can be examined in a functional framework by using the properties of harmonic functions. In particular, we get the following result:

\medskip
\nin {\bf Proposition 2 }[Gourieroux and Lee (2025)]:
In the nonlinear causal autoregressive representation of a mixed VAR(1) process, the dimension of under-identification in the functional space of nonlinear autoregressive models is finite and equal to $2 m$.

\medskip

\nin This result reveals the challenges and complexity of the identification of the nonlinear causal autoregression. 
%In comparison, the order of under-identification of a causal VAR model is $(m^2-m)/2$. 
The above under-identification is very large and exceeds the under-identification of linear causal VAR models. As pointed out in Section 5.1, it is possible to restrict the analysis to nonlinear autoregressions with Gaussian innovations, such that the components
$v_{1,t},...,v_{n,t}$  are independent N(0,1). However, this additional restriction is insufficient to solve the identification issue. Indeed, there exist nonlinear transformations of $v_t$, with respect to which the multivariate N(0, Id) distribution remains invariant [see Gourieroux and Lee (2025)], corresponding to "local matrix rotations".
%i.e. rotations depending on the level $v$.

\medskip

\nin {\bf Corollary 3: Functional multiplicity} 
There exists a functional multiplicity of innovation processes ($v_t$), which are cross-sectionally 
independent standard normal variables.

\medskip
It follows that the functional
identification of the  nonlinear autoregressive representation of a causal nonlinear innovation and nonlinear IRFs is a complicated issue in nonlinear autoregressive models, and in particular in the mixed VAR models. It resembles 
the identification issue of causal VAR and Structural (SVAR) models, which has remained unsolved for a long time. Until a better solution becomes available, we can
use one of the two following methods:

i) We can introduce identifying restrictions on function $a$ through parametric assumptions on the joint distribution of errors $\varepsilon_t$. For example, one could assume that this distribution is a non-Gaussian elliptical or t-Student. However, such a partial identifying assumption is difficult to justify due to the lack of a structural interpretation of errors
($\varepsilon_t$).

ii) A recursive structure of the model can be assumed, by analogy to the recursive structure of errors in causal VAR models that has been traditionally used for (S)VAR identification in the early macroeconomic literature [see Sims (1980) based on  Cholesky decomposition]. 

\subsection {Shock Ordering}

In practice one may prefer to adopt method ii). Note that 
since the observed process ($Y_t$) and the state process ($Z_t$) satisfy a one-to-one relationship, the state variables also satisfy a nonlinear autoregressive scheme:
\begin{equation}
Z_t = \tilde{a} (\underline{Z}_{t-1}, \tilde{v}_t),
\end{equation}
\nin where $\tilde{v}_t$ can be serially conditionally i.i.d. (and possibly restricted to be standard Gaussian). A recursive structure can be imposed on model (5.2) with the noncausal state variable to be shocked first under this approach (also referred to as "shock ordering"). This choice can be further motivated by the fact that in applied research,
the noncausal order is often equal to 1, i.e. $n_2 =1$ [see e.g. Hecq, Lieb and Telg (2016), Gourieroux and Jasiak (2017) and Section 6.3]. This implies that there is a common noncausal component generating the bubbles and local trends in $(Y_t)$. For example, in macroeconomic models, that noncausal component can be related to speculative bubbles in oil prices impacting jointly the price index, GDP and other macroeconomic variables. In financial applications, the single factor capturing the nonlinear dynamics can be interpreted as a systemic component, or a common bubble [Gourieroux and Zakoian (2017), Cubadda et al. (2023), Hall and Jasiak (2024)]. For a bivariate VAR(1) with $n_2=1$, we get an identifiable recursive form of the nonlinear autoregressive model (5.2):
\begin{eqnarray}
Z_{2,t} & = & G_2 ( Z_{2,t-1}; v_{2,t}),\\
Z_{1,t} & = & G_{1|2} (Z_{2,t}, Z_{1,t-1}; v_{1,t}),
\end{eqnarray}
\nin where function $G_2$ in $Z_{2,t} =  G_2 ( Z_{2,t-1}; v_{2,t})$ is the inverse of the function defining
the  nonlinear Gaussian causal innovation of $Z_2$, which is uniquely defined as:
\begin{equation}
v_{2,t}(Z) = \Phi^{-1}[F_2 ( Z_{2,t} | \underline{Y}_{t-1})] = \Phi^{-1}[F_2 ( Z_{2,t} | Z_{t-1})],
\end{equation}
\nin where $F_2$ is the conditional cumulative distribution function (c.d.f) of $Z_{2,t}$ given $Z_{t-1}$ and $\Phi$ is the c.d.f. of the standard Gaussian distribution. From equation (a.5) given in Appendix A.1, it follows that the conditional density of $Z_{2,t}$ given $Z_{t-1}$ has a closed form given by:
\begin{equation}
l(z_{2,t}| z_{t-1}) = \frac{l_2(z_{2,t})}{l_2(z_{2, t-1})} \; |det \, J_2| \; g_{\eta_2} (z_{2,t} - J_2 \, z_{2,t-1}),
\end{equation}
\nin where $g_{\eta_2}$ is the marginal density of $\eta_2$. The c.d.f. $F_2$ is the integral of 
$l(z_{2,t}| z_{t-1})$ over the set of admissible values of $z_{2,t}$.  By the Markov property of $(Z_{2,t})$ in Corollary 4 below, $Z_{1,t-1}$ does not appear in equation (5.6)
and we have $v_{2,t} = \Phi^{-1} [ F_2 (Z_{2,t}|Z_{2,t-1})]$.

Next, we append $v_{2,t}$ by the (Gaussian) ($v_{1,t}$) associated with the causal state variable(s) in this recursive model. In our illustration, we consider the case $n_1=1$ and define:
\begin{equation}
v_{1,t} (Z) =  \Phi^{-1}[F_{1|2} ( Z_{1,t} | Z_{2,t}, Z_{t-1})],
\end{equation}
\nin where $F_{1|2}$ is the conditional cumulative distribution function of $Z_{1,t}$ given $Z_{2,t}, Z_{t-1}$ \footnote{The system of equations (5.6)-(5.8) is the inverse of system (5.5)-(5.7) and it can be used for simulating the path of $Z_t$ from independent standard Gaussian drawing of $v_{1,t}, v_{2,t}$.}. From equation (a.5) given in Appendix A.1, we get the closed-form expression of the conditional density:
\begin{equation}
l(z_{1,t}| z_{2,t}, z_{t-1}) = l(z_t| z_{t-1})/l(z_{2,t}| z_{t-1}) = \frac{g_{\eta}
(z_{1,t} - J_1 z_{1, t-1}, \; z_{2,t} - J_2 z_{2, t-1} )}{g_{\eta_2} (z_{2,t} - J_2 z_{2, t-1})}.
\end{equation}
\nin The c.d.f. $F_{1|2}$ is the integral of 
$l(z_{1,t}| z_{2,t}, z_{t-1})$ over the set of admissible values of $z_{1,t}$.

When the noncausal state variable $Z_2$ is the first one to be shocked, it is easy to check that $v_{1,t}(Z)$ differs in general from the causal innovation $\eta_{1,t}$ associated with the latent causal component $Z_{1,t}$.
By applying the shock ordering, we are implementing the autoregressive model (5.2) in a recursive form (5.3)-(5.4). Moreover, we deduce from (5.5):

\medskip

\nin {\bf Corollary 4:} The noncausal state variable $(Z_{2,t})$ is a Markov process of order 1 with respect to the filtration associated with $(Y_t)$, [and also to the filtrations associated with $(Z_t)= (Z_{1,t}, Z_{2,t})$, and with $(Z_{2,t})]$.

\medskip

By construction, the structural noncausal shock $v_{2,t} (Z)$ is defined in a unique way (since $n_2=1$) and is a Gaussian white noise\footnote{See Gourieroux and Jasiak (2005) for the uniqueness of a nonlinear Gaussian innovation in univariate models.}. It is also independent of the multivariate shock $v_{1,t} (Z)$. Then, we can trace out nonlinear IRFs based on the effect of a change $\delta_2$ in $v_{2,t}$, with $v_{1,t}$ held constant. It is called henceforth the Common Bubble Shock (CBS) [See Online Appendix C for quantile estimation of $F_2 ( Z_{2,t} | Z_{2,t-1})$]. In the general case $n_2 = 1, n_1 \geq 1$, the associated multivariate impulse responses of $Z_{2,t}$ are identifiable, i.e. independent of other components ($v_{1,t}, v_{3,t},...,v_{n,t}$).

In practice, when there is a single noncausal state variable, the filtering algorithm of Section 4 can be completed by the two following steps:

\nin step 7. The approximations of nonlinear  causal innovations $\hat{v}_{2,t}(Z)$ can be computed from the estimated distributions as
$\hat{v}_{2,t}(Z) = \Phi^{-1}(\hat{F}_{2,T} (\hat{Z}_{2,t}| \hat{Z}_{2,t-1}))$ by applying the formula of  predictive density (5.6) with $l_2$ replaced by $\hat{l}_2$ and $g_{\eta_2}$ replaced by $\hat{g}_{\eta_2}$, i.e. the kernel-smoothed empirical density of $\hat{\eta}_{2,t}, \, t=1,...,T$.

\nin step 8. Next, $\hat{v}_{2,t}(Z)$ can be appended by the approximated nonlinear causal innovations $\hat{v}_{1,t}$, independent of $\hat{v}_{2,t}(Z)$, which are defined as: $\hat{v}_{1,t} = \Phi^{-1}(\hat{F}_{1|2,T} (\hat{Z}_{1,t}| \hat{Z}_{2,t}, \hat{Z}_{t-1})), \; \mbox{for} \;  n_1=1$.

\nin These causal innovations can be computed from the predictive density formula (5.8) with $g_{\eta_2}$ and $g_{\eta}$ replaced by their empirical counterparts.

\setcounter{equation}{0}\def\theequation{6.\arabic{equation}}
\section{Illustration}

This Section illustrates the nonlinear forecasts from the mixed VAR(1) model and the nonlinear IRF analysis. Subsection 6.1 presents a simulation study that examines the oos forecasts from the model. In subsection 6.2, the semi-parametric GCov estimators, nonlinear innovations and IRFs are  illustrated
in an application to 
a bivariate series of US GDP rates and oil prices. Additional simulations and empirical results are provided in Online Appendices D and E.

\subsection{Simulation Study}

We consider a simulated bivariate mixed VAR(1) process
with the following  matrix of autoregressive coefficient:
$\Phi = \left( \begin{array}{cc} 0.7 & -1.3 \\ 0 & 2 \end{array} \right),
$ with eigenvalues 0.7 and 2, located inside and outside the unit circle, respectively . The errors follow a bivariate noise with independent components both t-student distributed with  $\nu$ =4 degrees of freedom, mean zero, and variance equal to  $\nu/(\nu-2)=2$. The matrix A is as follows:
$A = \left( \begin{array}{cc} 1 & -1 \\ 0 & 1 \end{array} \right).
$ The simulated paths of the series of length 500 is displayed in Figure 1. The solid (black) line represents process $(Y_{1t})$ and the dashed (red) line represents process $(Y_{2t})$.
\nin The sequence of spikes in the noncausal component $Y_{2t}=Z_{2t}$ impacts the component $Y_{1t}$ through the recursive form of matrix $\Phi$.

Let us now consider the forecasts based on parameter estimates.
The Generalized Covariance (GCov) estimate of matrix $\Phi$ is obtained by minimizing
the portmanteau statistic computed from the auto- and cross-correlations up to and including lag  $H=10$
of the errors  $\varepsilon_t = Y_t - \Phi_1 Y_{t-1}$
and their squared values [see Gourieroux and Jasiak (2023)]\footnote{The finite sample properties of the GCov estimator are illustrated in Gourieroux and Jasiak (2023).}.
The estimated autoregressive matrix is 
$\hat{\Phi} = \left( \begin{array}{cc} 0.724 & -1.452 \\ -0.030 & 1.993 \end{array} \right),$
with eigenvalues $\hat{\lambda}_1=0.690$, $\hat{\lambda}_2=2.027$, which are close to the
true values $\lambda_1 = J_1 = 0.7$ and $\lambda_2 = J_2 = 2.0$. The standard errors of $\hat{\Phi}$ obtained by bootstrap
are 0.023, 0.308, for the elements of the first row, and  0.009, 0.120 for the
elements of the second row.

\begin{center}
\begin{figure}
\centering
\includegraphics[width=12cm,height =3cm]{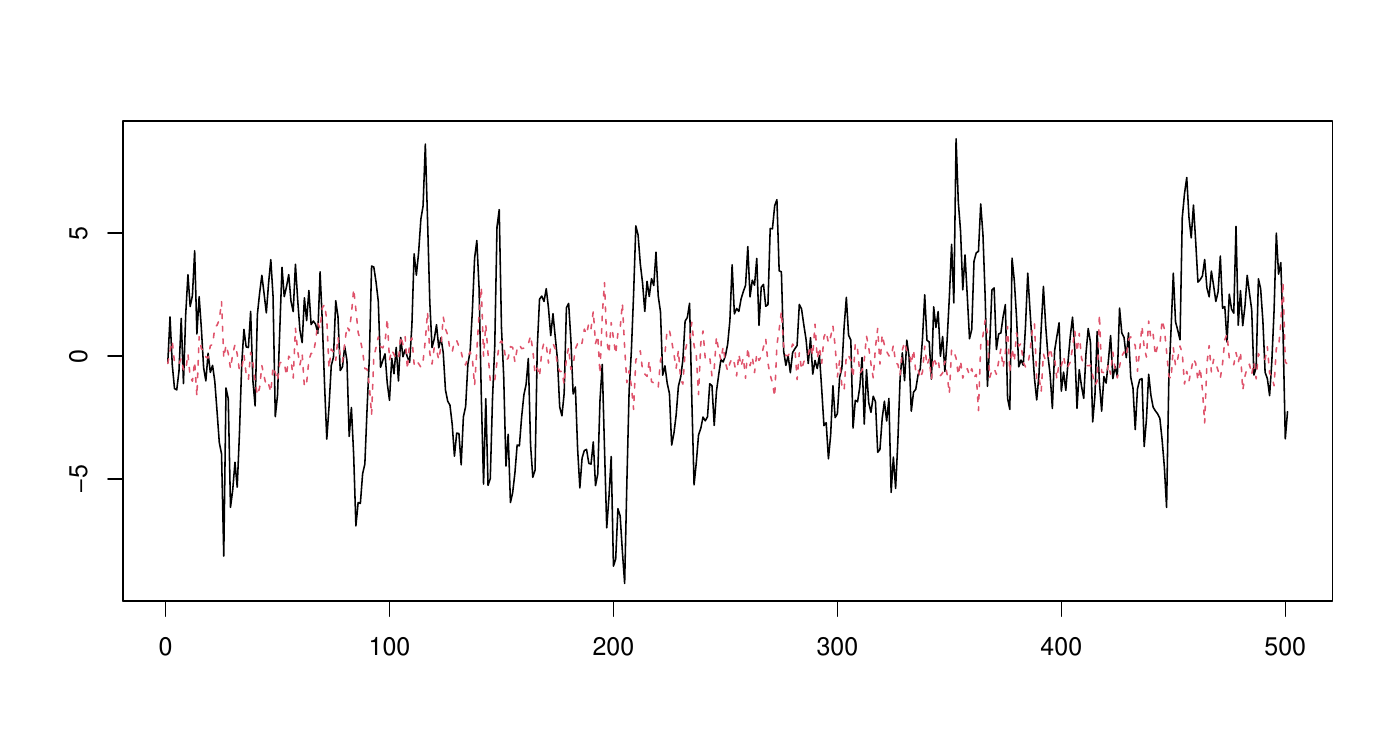}
\caption{Bivariate Mixed VAR(1) Process: $Y_1$: solid line, $Y_2$: dashed line}
\end{figure}
\end{center}

\vspace{-1cm}
\nin After estimating $\Phi$, the GCov estimated errors $\hat{\varepsilon}_t=Y_t - \hat{\Phi} Y_{t-1}$ are computed. Matrices $A$ and $A^{-1}$ are identified from the real Jordan representation of matrix $\Phi$, up to scale factors. The estimated matrix $\hat{A}^{-1}$ computed from the normalized Jordan decomposition of $\hat{\Phi}$ is:

$ \hat{A}^{-1} =  \left( \begin{array}{cc} 0.022 & 0.025 \\ -0.022 & 0.974 \end{array} \right)$. It corresponds to the true matrix $A^{-1} =  \left( \begin{array}{cc} 1 & 1 \\ 0 & 1 \end{array} \right)$ up to scale factors of about 0.02 and 0.97 for each column.
We use these estimates to approximate the causal and noncausal components displayed in Figure 2.

Let us now consider the oos nonlinear forecast one step ahead performed at date $T=500$ when the process takes values  $Y_{1, 500} =-3.367$ and $Y_{2, 500}=-0.239$. The true values of $Y_{1,501}$ and $Y_{2, 501}$ 
 are -2.260 and  -0.331, respectively. The nonlinear forecasts are summarized by the predictive density, whose mode provides pointwise predictions of $Y_{1, T+1}$ and
$Y_{2, T+1}$. It is estimated from formula (3.2) one-step ahead oos by using a kernel estimator over a grid of 100 values below and above $Y_1$ and $Y_2$, with Gaussian kernels and bandwidths $h_2=1$
and $h_{11} = s.d.(\varepsilon_1), h_{12} = s.d.(\varepsilon_2)$ (see Online Appendix B for kernel density estimators).
The estimated point forecasts are $\hat{Y}_{1,501}= -2.80$ and  $\hat{Y}_{2,501}= -0.30$. The estimated prediction intervals at level 0.80 determined from the predictive density are  [-4.80, -0.80] 
for $Y_{1,501}$ and [-2.60, 2.10]
 for $Y_{2,501}$. The rationale for choosing level 80\% is to ensure a sufficiently large number of observations in the tails for reliable estimation of the quantiles of predictive density. Both prediction intervals contain the true future values of the process. Additional results on the coverage of the estimated prediction interval and on the estimated prediction set uncertainty are provided in Online Appendix D.

\begin{center}
\begin{figure}
\centering
\includegraphics[width=12cm,height =3cm, angle = 0]{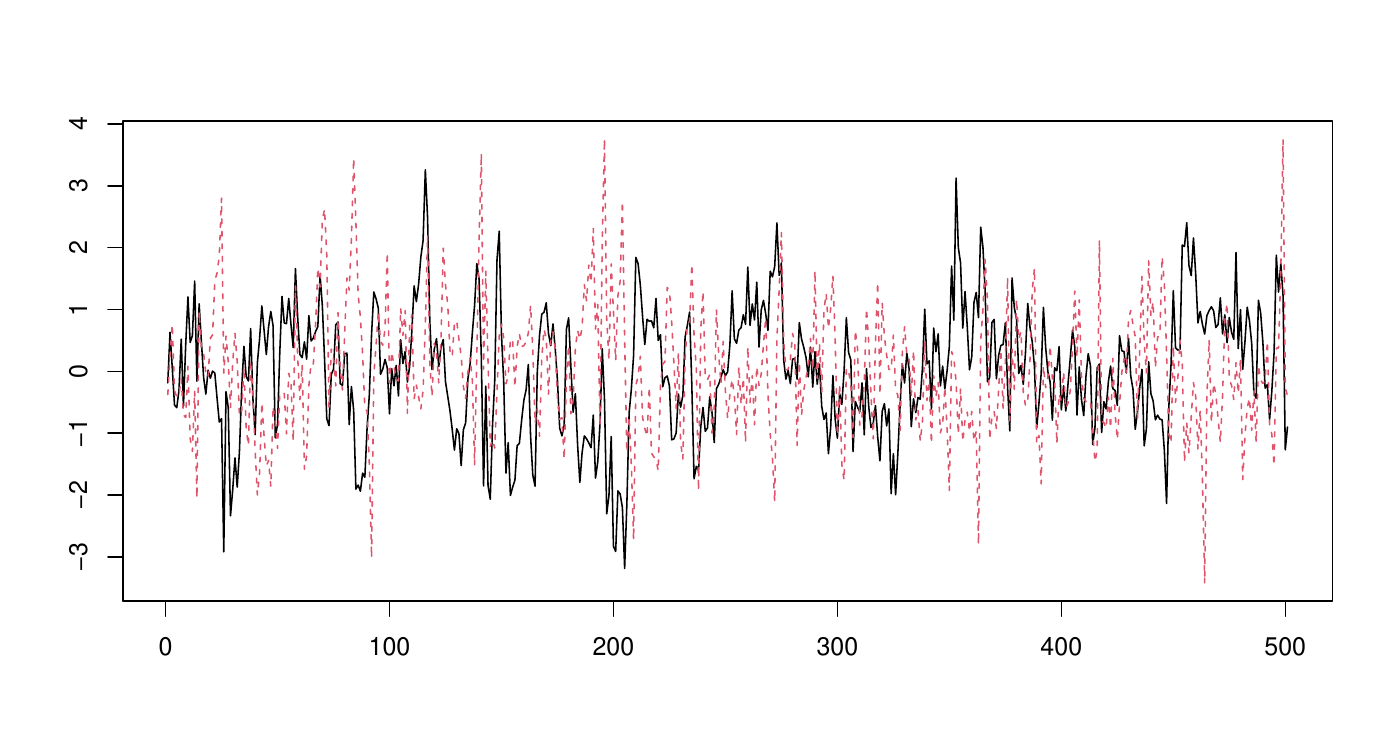}
\caption{State variables: $\hat{Z}_1$: solid line, $\hat{Z}_2$: dashed line}
\end{figure}
\end{center}

\vspace{-2cm}

\subsection{Application to Real Oil Prices and Real GDP Growth Rates}

In this Section, we apply the mixed VAR(1) model to analyse jointly the real oil prices and the real GDP growth rates.

\nin {\bf a) The data}

We examine the quarterly series of oil prices and US GDP growth rates over the 
period: Q1 1986 - Q2 2019. The real US GDP growth rate series is calculated from  Real Gross Domestic Product,
%Billions of Chained 2012 Dollars%,
Quarterly, Seasonally  Adjusted Annual Rate available at
at https://fred.stlouisfed.org from the Federal Reserve Economic Data.%\footnote{Other measures of global real economic activity could be used, especially to distinguish the global business cycle fluctuations from fluctuations on the global commodity markets [See, Kilian (2009), (2019), Kilian, Zhou (2018), Hamilton (2018) for the indexes of real economic activity].}
	
The oil prices are provided online by the US Energy Information Administration
under the Short-Term Energy Outlook Real and Nominal Prices, March 2023 \footnote{ called 
Quarterly Average Imported Crude Oil Price/barrel, Real Price, deflated by the US Consumer Price CPI index.}. This series approximates "the price of oil paid by US refiners for crude oil purchased from abroad" examined by Kilian and Vigfusson (2017).
The series of GDP rates and oil prices (divided by 10) of length T=134 are displayed in Figure 3. We observe, for example, in year 2008, a pronounced bubble in oil prices with a strong negative impact on the GDP growth rate.
 
\nin {\bf b) Estimation} 

We estimate the causal-noncausal VAR(1) from the demeaned series of GDP growth rates (series 1) and demeaned oil prices divided by 10 (series 2). The sample mean of growth rates is 0.640 and the mean of rescaled oil prices is 6.348.  The estimated autoregressive matrix is
$\hat{\Phi} = \left[ \begin{array}{cc} 0.2987 & -0.0229 \\ -0.1527 & 1.0690 \end{array} \right]$
with standard deviations of autoregressive coefficients of 0.014, 0.001, 0.032, 0.004, respectively. The eigenvalues are 0.294 and 1.073. The densities of estimated errors provided in Online Appendix E, Figure a.5 are non-Gaussian.

The presence of a noncausal root reflects the nonlinear dynamic features %corresponding to the sequence of speculative bubbles 
such as the spikes and a bubble observed
in Figure 3.  The spikes and bubble in oil prices are accommodated by the strictly stationary mixed VAR(1) model of GDP rates and oil price levels. 
%The mixed causal-noncausal model is an alternative to the threshold autoregressive model of the series of GDP rates and first differences of oil price logarithms\footnote{The analysis of oil prices is often performed on diff-logs. That alternative approach could lead to over-differencing if the trend (bubble) is  a local explosive pattern in a strictly stationary process.} [see e.g. Herrera, Lagalo and Wada (2015), Kilia and, Vigfusson (2017)] discussed at the end of this section and in On-line Appendix E.

\nin {\bf c) Prediction}

We perform out-of-sample (oos) predictions from the mixed VAR(1) at the points indicated in Figure 3 below, which mark the bubble episode at the end of the sample to ensure a sufficiently large number of prior observations for estimation.

\begin{figure}[h]
\centering
\includegraphics[width=12cm,height =3cm, angle = 0]{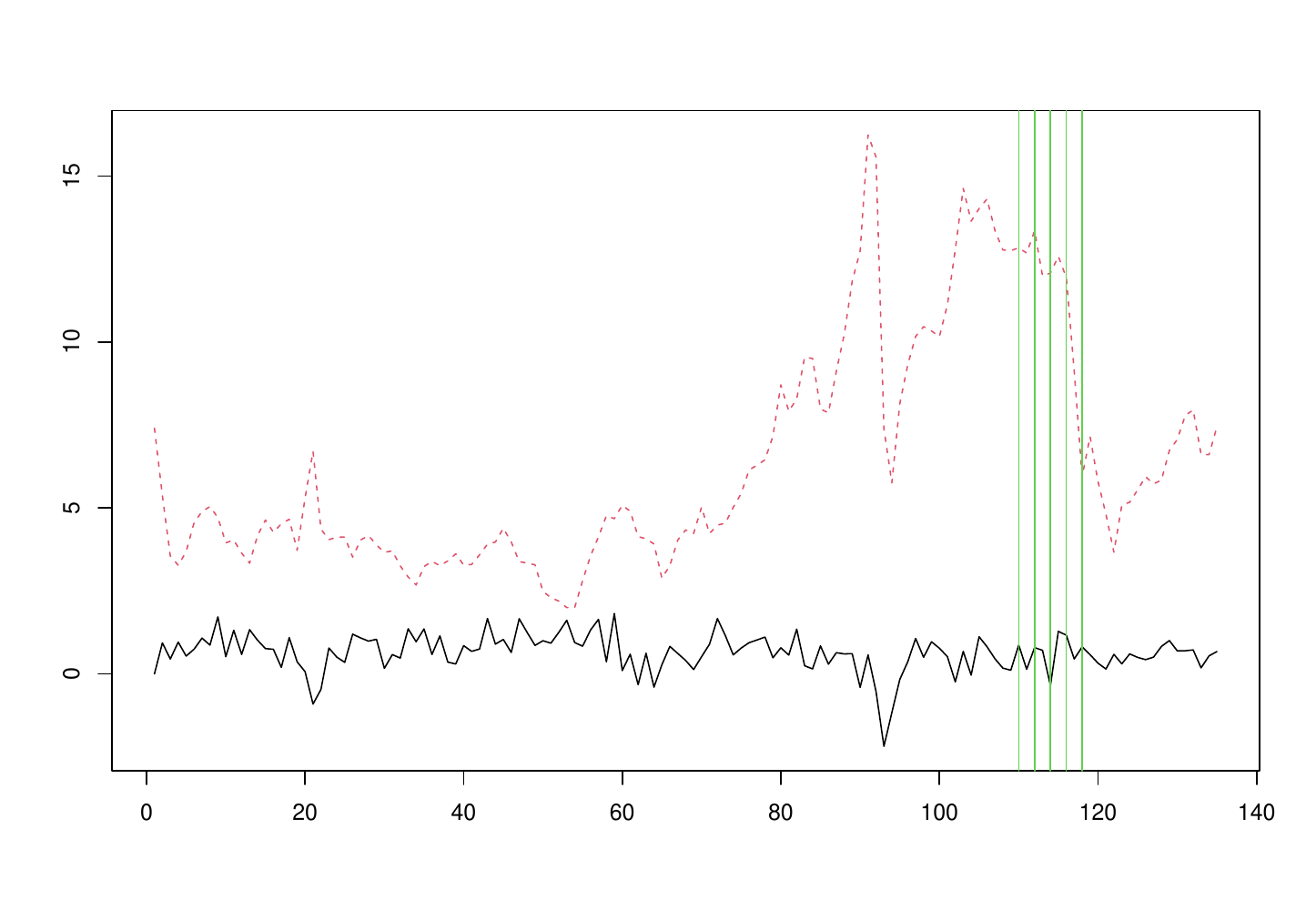}
\caption{Quarterly Real GDP Rate (black) and Oil Price (red), Q1 1986 -Q2 2019}
\end{figure}

During the selected period, both series displayed several small sudden changes, before the oil price dropped.
%More specifically, the oil prices at times T=110, 112, 114, 115, 116 and 118 were 126.84, 119.98, 125.86, 119.44, 91.14 and 71.52. The growth rate spiked briefly and took the values: 0.13, 0.71, 1.28, 1.16, 0.44 and 0.58. These changes are reflected in the values of demeaned growth rates and of rescaled and demeaned oil prices. 
%From the technical point of view, we chose a period when a fair number of observations is available at the end of the trajectory to evaluate the predictive density reasonably well.
The predictive densities are evaluated at each point over a grid of 200 points, equidistant by 0.1 below and above the last observed value of each variable.
We use again Gaussian kernels and bandwidths $h_2=s.d.(Z_{2})$,
$h_{11} = s.d.(\varepsilon_1)$, and $h_{12} = s.d.(\varepsilon_2)$ to estimate one-step ahead oos predictive density (3.2) [See Online Appendix E, Figures a.7-a.11 for predictive density plots which vary in $T$\footnote{because of the increasing information set and the dependence on the current environment of $Y_T$. } and become bi-modal on-bubble when the probability of crash increases].

%\singlespacing

\begin{table}
\begin{center}
\begin{tabular}{|c|c|c|c|c|c|c|c|c|}
\hline
T+1 & \multicolumn{2}{|c|}{forecast}& \multicolumn{2}{|c|}{interval growth}& \multicolumn{2}{|c|}{interval oil}& \multicolumn{2}{|c|}{true values}\\
\hline
& growth & oil & q11 & q12 & q21 & q22 & growth & oil \\
\hline
111 & -0.037 &  6.937 &  -1.437 & 0.962 & 4.337 & 8.537 & -0.501 & 6.335 \\
113 & -0.015 & 7.333  &  -1.315 &  0.784 & 4.933 & 8.833 & 0.070 &  5.649 \\
115 & -0.359 & 6.221  & -1.559  &  0.540 &  3.921  & 7.921 & 0.643 &  6.237 \\
116 & 0.145 & 6.451 &  -1.054 & 1.045 & 4.051 & 8.251 &  0.524 & 5.595 \\
117 & 0.121 & 5.862 & -1.078 &   1.021 &  3.562 & 7.562 & -0.191 & 2.765 \\
119 & 0.077 & -0.379 & -0.922 & 1.077 & -2.479 &  1.420 & -0.060 & 0.803 \\
\hline
\end{tabular}
\caption{On-bubble Prediction Intervals at 80\% conditioned on data at $T=110:118$\\
The predictions (col.2)  and prediction intervals (col.3:4) computed oos one-step ahead at selected dates on-bubble for the future values at times T+1 given in col.1}
\end{center}

\end{table}

Table 1 reports the point forecasts obtained from the main modes of predictive densities, and the forecast intervals obtained from the sample quantiles of estimated predictive densities.
The forecast interval is at level 80\% to ensure a sufficient number of observations to estimate the quantiles of the predictive density and to alleviate the potential effect of bimodality. We also consider the predictions of the last 2 points in the trajectory. The results in Table 2 are computed according to the procedure described earlier, starting from
the conditioning time T=132 when growth rate $Y_{1, 132}=-0.458$ and oil $Y_{2, 132}=0.268$ are closer to their averages.

%\singlespacing
\begin{table}[h]

\begin{center}

\begin{tabular}{|c|c|c|c|c|c|c|c|c|}
\hline
T+1 & \multicolumn{2}{|c|}{forecast}& \multicolumn{2}{|c|}{interval growth}& \multicolumn{2}{|c|}{interval oil}& \multicolumn{2}{|c|}{true values}\\
\hline
& growth & oil & q11 & q12 & q21 & q22 & growth & oil \\
\hline
133 & -0.099 & 0.340 & -1.099 &  0.800  & -1.759 & 2.140 & -0.097 & 0.257 \\
134 & -0.020 & 0.307 & -1.020 & 0.879  & -1.792  & 2.007 &   0.032  &   1.103 \\
\hline
\end{tabular}
\caption{Off-bubble Prediction Intervals at 80\%, $T=132 : 133$ \\
The predictions (col.2)  and prediction intervals (col.3:4) computed oos one-step ahead off-bubble for the future values at times T+1 given in col.1}
\end{center}
\end{table}
%\doublespacing

\nin These prediction intervals are conditional and depend on the date and distance in time from a bubble. 
%We expect asymmetries in the predictive densities during a bubble increase and decrease.
We observe that the prediction intervals on-bubble are longer than off-bubble.
In addition,
we use the approach outlined in Section 4 to estimate the confidence set of the
prediction interval for $Y_{1,134}$. It is based on 100 backcast replications of the series conditional on $Y_T$. The confidence set of prediction interval for the demeaned growth rate at level $1-\alpha_2 =0.95$ is: $\widehat{CSPI}(y, \alpha_1, \alpha_2) = [-1.546, 1.405]$. This interval is at a level higher than the prediction intervals in Table 2, accounts also for the estimation risk and is longer.

\nin {\bf d) Bubble shock}

Let us now examine the effect of a set of the following values of the CBS: $\delta_2= -2, -1, 0, 1, 2$ to $z_{2,T}$  performed at date $T=110$ during the bubble episode and at time $T=133$ after the bubble, under the identifying assumption of a recursive form (5.3)-(5.4). Since $Z_t= A^{-1} Y_t$, that shock can originate from either the oil prices, or growth rates, or both. 
\begin{figure}[h]
\centering
 \centering
  \begin{subfigure}[b]{0.4\linewidth}
  \centering
\includegraphics[width=\linewidth,height =2cm,]{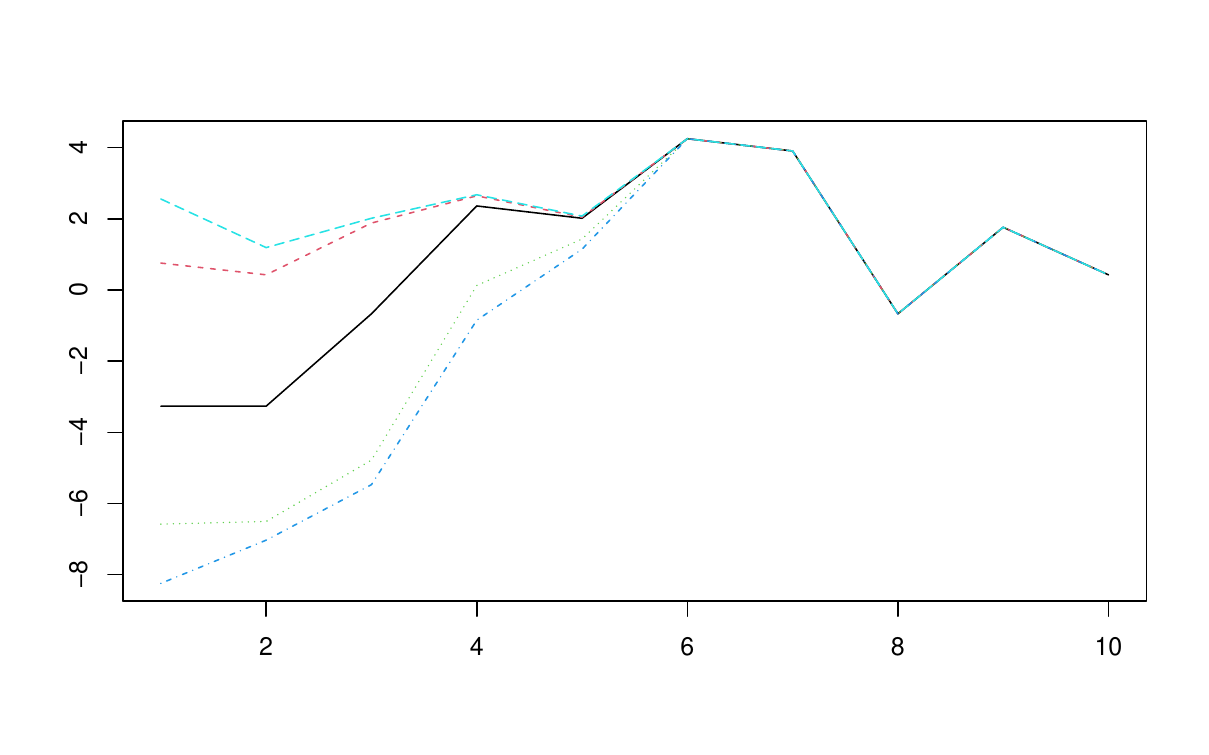}
     \caption{on bubble T=110}
  \end{subfigure}
  \centering
  \begin{subfigure}[b]{0.4\linewidth}
  \centering
\includegraphics[width=\linewidth,height =2cm,]{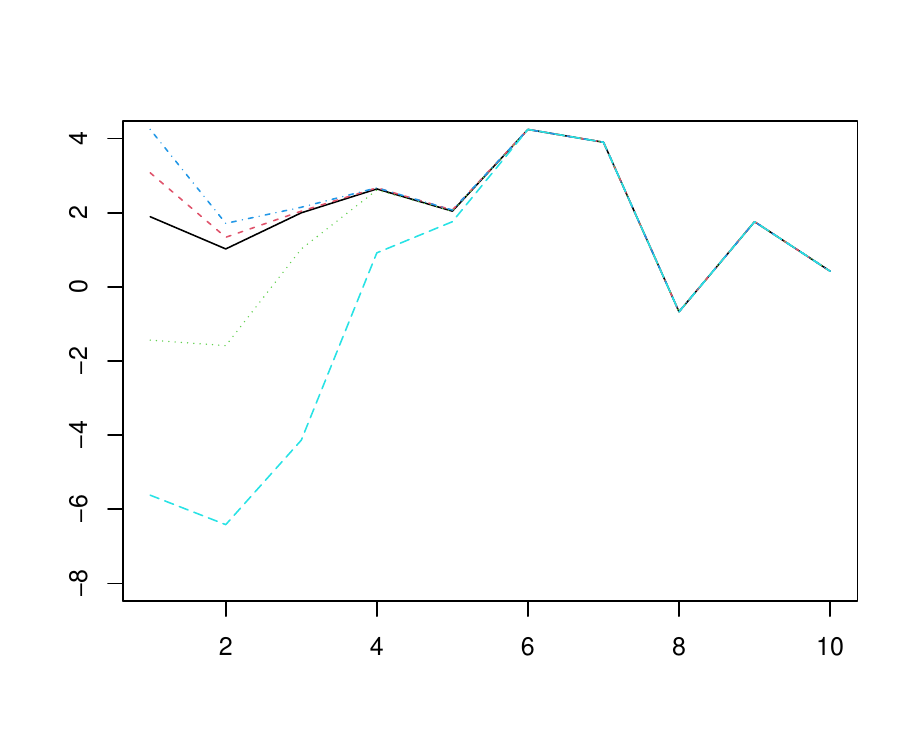}
    \caption{off bubble, T=133}
  \end{subfigure}
\caption{Shocks to $Z_2$, black line: baseline $\delta=0$, colored: response functions:
$\delta=+1$: red (thin dashed), $\delta=+2$: bright blue (thick dashed), $\delta=-1$: green (dotted) $\delta=-2$ dark blue: (dash-dotted)}
\end{figure}

The impulse responses in Figure 4 are computed conditionally on the path of the process up to and including date $T$. Because the dynamic model is nonlinear, the IRF is nonlinear in $\delta_2$, and the shock effects need to be compared with a baseline. 
The baseline path of $Z_{2, T+1}^b,...,Z_{2, T+10}^b$ is computed from 10 random values of standard Normal used as the future $v_{2, T+1},...,v_{2, T+10}$. Next, a shock $\delta$ is added to $v_{2, T+1}$ and the consecutive values of shocked $Z_{2, T+1}^{\delta},...,Z_{2, T+10}^{\delta}$ are calculated recursively by inverting the conditional c.d.f. [See Online Appendix C]. 

To interpret the shocks, we compare the relative size of shock effects to the baseline, and we conclude that a shock has dissipated when the shocked path overlaps with the baseline. We observe that, on-bubble, shocks to $Z_2$ are more long-lasting and more symmetric around the baseline. Off-bubble, we observe that a shock of $\delta=-2$ dissipates much slower than other shocks. 
Next, we perform shocks of the same size to the causal component $Z_1$ using a similar approach (Figure 5). We find that shocks to $Z_1$ have weaker effects and dissipate quickly, which is consistent with the fact that the bubble is driven mainly by the noncausal component $Z_2$. 

\begin{figure}
\centering
 \centering
  \begin{subfigure}[b]{0.4\linewidth}
  \centering
\includegraphics[width=\linewidth,height =2cm, angle = 0]{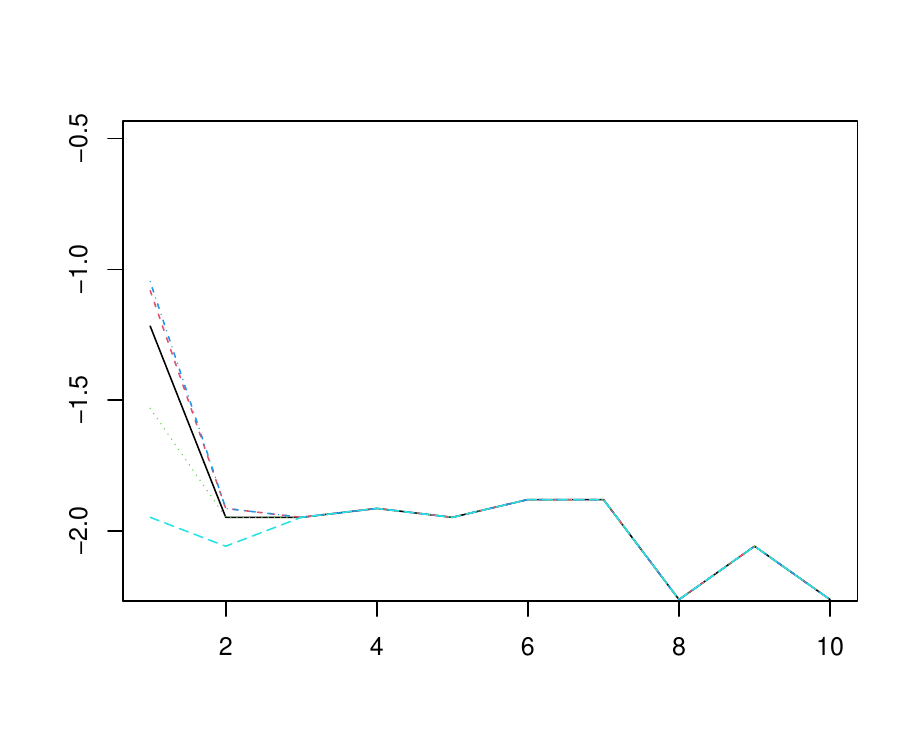}
\caption{on bubble, T=110}
 \end{subfigure}
  \centering
  \begin{subfigure}[b]{0.4\linewidth}
  \centering
\includegraphics[width=\linewidth,height =2cm, angle = 0]{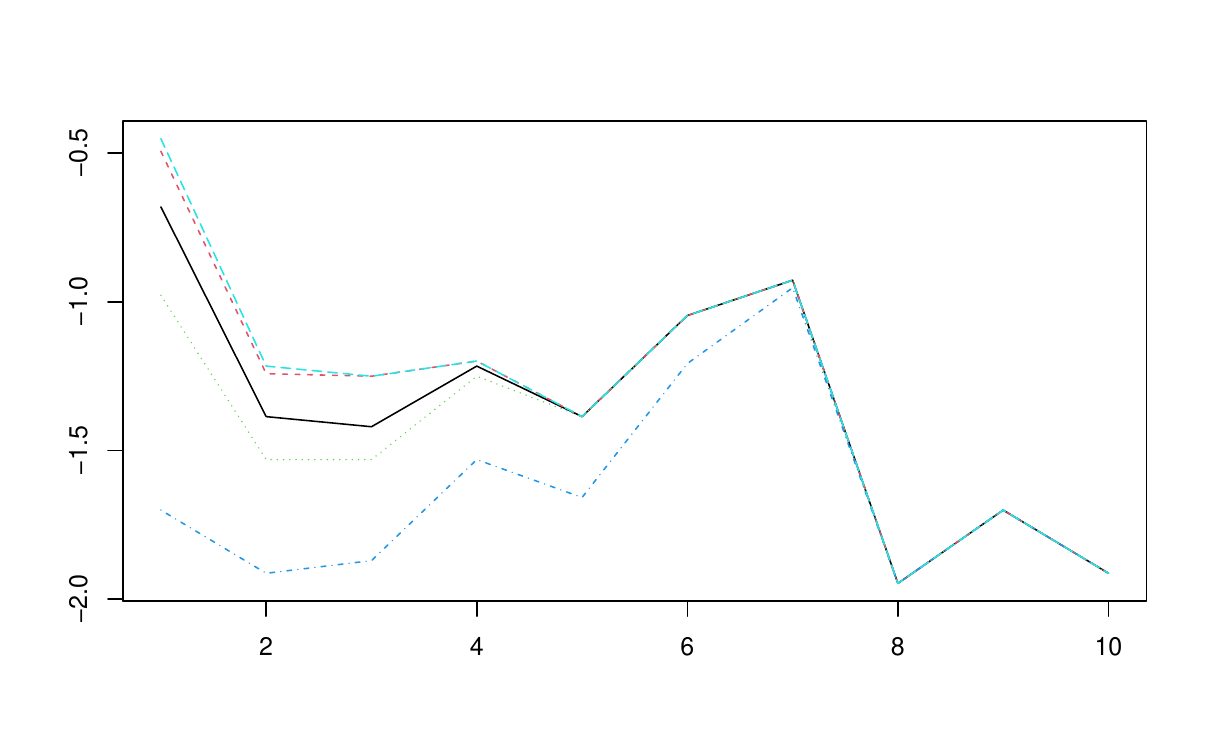}
\caption{off bubble, T=133}
  \end{subfigure}
\caption{Shocks to $Z_1$, black line: baseline $\delta=0$, colored: response functions:
$\delta=+1$: red (thin dashed), $\delta=+2$: bright blue (thick dashed), $\delta=-1$: green (dotted) $\delta=-2$ dark blue: (dash-dotted)}
\end{figure}

The shocked component $Z_2$ can be combined with the values of causal component $Z_1$ conditional on its own past and the shocked values of $Z_2$. This is done by predicting $Z_1$ from the mode of the conditional density $l(z_{1,t}|z_{2,t}, z_{1, t-1})$ given in (5.8) and kernel estimated. Then, we can approximate the shocks to $y_1$ and $y_2$ as $Y^{\delta} = A Z^{\delta}$. For illustration, Figure 6 displays the shocked values of 
$Y_{2, T+1}^b,...,Y_{2, T+10}^b$.

\begin{figure}
\centering
 \centering
  \begin{subfigure}[b]{0.4\linewidth}
  \centering
\includegraphics[width=\linewidth,height =2cm]{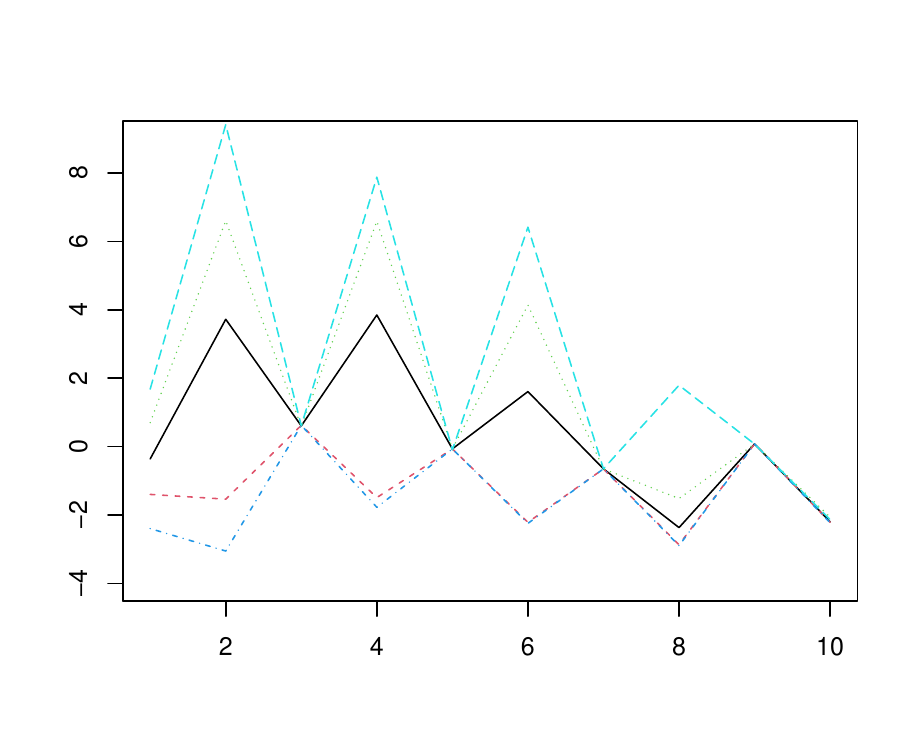}
\caption{on bubble, T=110}
  \end{subfigure}
  \centering
  \begin{subfigure}[b]{0.4\linewidth}
  \centering
\includegraphics[width=\linewidth,height =2cm]{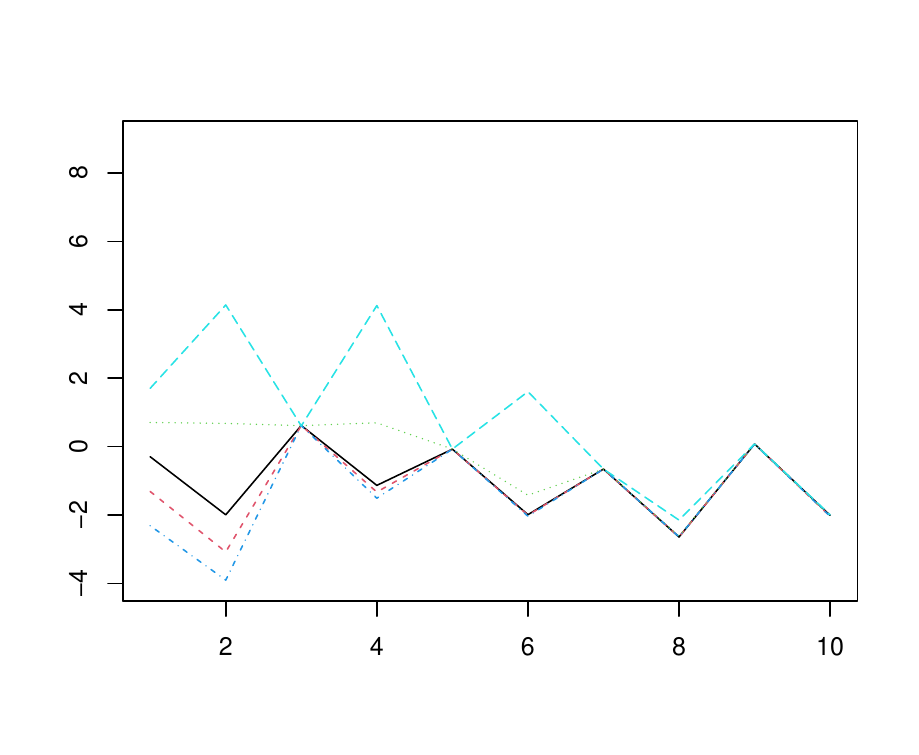}
\caption{off bubble, T=133}
  \end{subfigure}
\caption{Shocks to $y_2$, black line: baseline $\delta=0$, colored: response functions:
$\delta=+1$: red (thin dashed), $\delta=+2$: bright blue (thick dashed), $\delta=-1$: green (dotted) $\delta=-2$ dark blue: (dash-dotted)}
\end{figure}

We find that  $Y_2$ is more responsive to on-bubble shocks. Off-bubble, the shocks dissipate faster except for the large positive shock, which again takes longer to disappear.

\section{Concluding Remarks}

This paper considers  oos nonlinear forecasting and backcasting in a mixed VAR model. It introduces a closed-form expression of forward (resp. backward) predictive density for forecasting (resp. backcasting) from mixed VAR models. As a post-estimation inference method, we adjust the estimated prediction interval by a conditional "backward" bootstrap and introduce the confidence set of the estimated prediction interval. 
%It allows for an assessment of the estimation effect on the forecast. A backcasting enhanced CI  allows also for increasing the level of a quantile-based prediction interval. This is beneficial when the quantiles are poorly estimated because of scarce observations on the tails or an insufficient number of points at which the predictive density is evaluated     
A definition of causal (past-dependent) nonlinear innovations for mixed VAR models is also given.
Since the causal nonlinear innovations are not uniquely defined, their identification is examined and the IRF analysis following a shock to the noncausal component is discussed.
%in detail.  Moreover, we introduce a filtering algorithm for in-sample inference.

%The approach is illustrated by a simulation study highlighting the differences between the errors, residuals and innovations of a mixed (S)VAR model. 
For illustration, the proposed approach is applied to the analysis of the joint dynamics of  a bivariate series of oil prices and real GDP growth rates. We find that the  noncausal state variable captures the explosive patterns, including bubbles and spikes. We examine the IRFs and observe different effects of a shock to the noncausal component on- and off-bubble. The mixed VAR models offers a parsimonious representation of nonlinear dynamic processes and allow for advanced analysis of the dynamics during a bubble episode.

\medskip
\singlespacing
\nin {\bf Funding:} This work was supported by the Natural Sciences and Engineering Research Council of Canada (NSERC).

\begin{center}
\textbf{R E F E R E N C E S}
\end{center}

\nin Beresteanu, A., Molchanov, I., and F. Molinari (2017): "Partial Identification Using Random Set Theory", Journal of Econometrics, 16, 17-32.

\nin Blasques, F.,  Koopman, S., Mingoli, G. and S., Telg (2025): "A Novel Test for the Presence of Local Explosive Dynamics", Journal of Time Series Analysis, forthcoming. 

\nin Breidt, F., Davis, R., Lii, K., and M., Rosenblatt (1991)~: "Maximum Likelihood Estimation for Noncausal Autoregressive
Processes", Journal of Multivariate Analysis, 36, 175-198.

\nin Cambanis, S., and I., Fakhre-Zakeri (1994): "On Prediction of Heavy-Tailed Autoregressive Sequences Forward versus Reversed Time", Theory of Probability and its Applications, 39, 217-233. 

\nin Cavaliere, G., Nielsen, H., and A., Rahbek (2020): "Bootstrapping Noncausal Autoregressions with Applications to Explosive Bubbles Modelling", Journal of Business and Economic Statistics, 38, 55-67.

%\nin Comon, P. (1994): "Independent Component Analysis: A New Concept?", Signal Process., 36, 287-314. \vspace{1em}

\nin Cubadda, G., Hecq, A.,  and S., Telg (2019): "Detecting Co-Movements in Non-Causal Time Series, Oxford Bulletin on Economics and Statistics, 697-715.

\nin Cubadda, G., Hecq, A.,  and E., Voisin (2023): "Detecting Common Bubbles in Multivariate Mixed Causal-Noncausal Models", Econometrics, 11, 9.

\nin Cubadda, G., Giancaterini, F., Hecq, A. and J. Jasiak (2024): "Optimization of the Generalized Covariance Estimator in Noncausal Processes",  Statistics and Computing, 34, 127.

\nin Davis, R., and L., Song (2020)~: "Noncausal Vector AR Processes with Application to Economic Time Series", Journal of Econometrics, 216, 246-267.

\nin De Truchis, G., Freis, S. and A., Thomas (2025): "Forecasting Extreme Trajectories Using Semi-Norm Representations", working paper Paris Dauphine University.

\nin Fries, S., and J.M., Zakoian (2019): "Mixed Causal-Noncausal Autoregressive Processes", Econometric Theory, 35, 1234-1270.

\nin Gelfand, A., and A., Smith (1992): "Bayesian Statistics without Tears: A Sampling-Resampling Perspective", Annals of Statistics, 46, 84-88. 

\nin Giancaterini, F., Hecq, A., Jasiak, J. and A. Manafi-Neyazi (2025): "Regularized Generalized Covariance Estimator", ArXiv 6390395.

\nin Gonzalves, S., Herrera, A., Kilian, L., and E., Pesavento (2021): 'Impulse Response Analysis for Structural Dynamic Models with Nonlinear Regressors", Journal of Econometrics, 225, 107-130. 

\nin Gourieroux, C., and A., Hencic (2015) : "Noncausal Autoregressive Model in Application to Bitcoin/USD Exchange Rates", Econometrics of Risk, Studies in Computational Intelligence, 583:17-40

\nin Gourieroux, C., and J., Jasiak (2005)~: "Nonlinear Innovations and Impulse Responses with Application to VaR Sensitivity", Annals of Economics and Statistics, 78, 1-31.

\nin Gourieroux, C., and J., Jasiak (2016): "Filtering, Prediction, and Simulation Methods for Noncausal Processes", Journal of Time Series Analysis, 37, 405-430.

\nin Gourieroux, C., and J., Jasiak (2017): "Noncausal Vector Autoregressive Process: Representation, Identification and Semi-Parametric Estimation", Journal of Econometrics, 200, 118-134. 

\nin Gourieroux, C., and J., Jasiak (2023): "Generalized Covariance Estimator", Journal of Business and Economic Statistics, 41, 1315 -1357. 

\nin Gourieroux, C., Jasiak, J., and M., Tong (2021): "Convolution-Based Filtering and Forecasting: An Application to WTI Crude Oil Prices", Journal of Forecasting, 40, 1230-1244. 

\nin Gourieroux, C., and Q., Lee (2025): "Identification and Impulse Response Functions for Nonlinear Dynamic Models", ArXiv 2506.13531.

\nin Gourieroux, C., Monfort, A., and J.P., Renne (2017)~: "Statistical Inference for Independent Component Analysis", Application to Structural VAR Models", Journal of Econometrics, 196, 111-126.

%\nin Gourieroux, C., Monfort A., and J.P., Renne (2020): "Identification and Estimation in Nonfundamental Structural VARMA Models", Review of Economic Studies, 87, 1915-1953.
%\vspace{1em}

\nin Gourieroux, C., and J.M., Zakoian (2017)~: "Local Explosion Modelling by Noncausal Process", Journal of the Royal Statistical Society, B, 79, 737-756.

%\nin Guay, A. (2021): "Identification of Structural Vector Autoregressions Through Higher Unconditional Moments", Journal of Econometrics, 225, 27-45. \vspace{1em}

\nin Hall, M., and J., Jasiak (2024): "Modelling Common Bubbles in Cryptocurrency Prices", Economic Modelling, 139, 106782.

%\nin Hamilton, J. (2018)~: "Measuring Global Economic Activity", DP University of California at San Diego. 

\nin Hecq, A. , Lieb, L. and S., Telg (2016): "Identification of Mixed Causal-Noncausal Models in Finite Samples", Annals of Economics and Statistics, 123/124, 307-331. 

\nin Herrera, A., Lagalo, L., and T., Wada (2015)~: "Asymmetries in the Response of Economic Activity to Oil Price Increases and Decreases ?", Journal of International Money and Finance, 50, 108-133.

\nin Imbens, G., and C., Manski (2004): "Confidence Intervals for Partially Identified Parameters", Econometrica, 72, 1845-1857. 

%\nin Kilian, L. (2009)~: "Not Oil Price Shocks are Alike : Disentangling Demand and Supply Shocks in the Crude Oil Market", American Economic Review, 99, 1053-1069.

%\nin Kilian, L. (2019)~: "Measuring Global Real Economic Activity : Do Recent Critiques Hold up to Scrutiny ?", Economics Letters, 178, 106-110. 

\nin Kilian, L., and R., Vigfusson (2017)~: "The Role of US Oil Price Shocks in Causing US Recessions", Journal of Money, Credit and Banking, 40, 1747-1776.

%\nin Kilian, L., and X., Zhou (2018)~: "Modeling Fluctuations in the Global Demand for Commodities", Journal International Money and Finance, 88, 54-78. 

\nin Lanne, M., and  J., Luoto (2016): "Noncausal Bayesian Vector Autoregression", Journal of Applied Econometrics, 31, 1392-1406. 

\nin Lanne, M., and P., Saikkonen (2011)~: "Noncausal Autoregressions for Economic Time Series", Journal of Time Series Econometrics, 3, 1-39.

\nin Lanne, M., and P., Saikkonen (2013)~: "Noncausal Vector Autoregression", Econometric Theory, 29, 447-481.

\nin Lof, M., and H., Nyberg (2017): "Noncausality and the Commodity Currency Hypothesis", Energy Economics, 65, 424-433.

\nin Molchanov, I., and F., Molinari (2018): "Random Sets in Econometrics", Econometric Society Monographs, Cambridge University Press.

\nin Nyberg, H., and P., Saikkonen (2014): 'Forecasting with a Noncausal VAR Model", Computational Statistics and Data Analysis, 76, 536-555. 

\nin Perko, L. (2001): "Differential Equations and Dynamical Systems", Springer, New York.

\nin Rosenblatt, M. (1952)~: "Remarks on Multivariate Transformations", Annals of Mathematical Statistics, 23, 470-472.

\nin Rosenblatt, M. (2012)~: "Gaussian and Non-Gaussian Linear Time Series and Random Fields", Springer Verlag.

\nin Sims, C. (1980): "Macroeconomics and Reality", Econometrica, 48, 1-48.

\nin Swensen, A. (2022): "On Causal and Non-Causal Cointegrated Vector Autoregressive Time Series", Journal of Time Series Analysis, 42, 178-196. 

\nin Tanner, M. (1993): "Tools for Statistical Inference", Springer Series in Statistics, 2nd edition, Springer, New York.

\nin Twumasi, C., and J., Twumasi (2022): "Machine Learning Algorithms for Forecasting and Backcasting
Blood Demand Data with Missing Values and Outliers: A Study
of Tema General Hospital of Ghana", International Journal of Forecasting, 38, 1258-1277.

\nin Velasco, C. (2023): "Identification and Estimation of Structural VARMA Models Using Higher Order Dynamics", Journal of Business and Economic Statistics, 41, 819-832. 

\nin Velasco, C., and I., Lobato (2018): "Frequency Domain Minimum Distance Inference for Possibly Noninvertible and Noncausal ARMA Models", Annals of Statistics, 46, 555-579. 

\doublespacing

\setcounter{equation}{0}\def\theequation{a.\arabic{equation}}
\section*{APPENDIX A} 

\begin{center}
{\bf APPENDIX A.1: Proof of Proposition 1}
\end{center}

\nin The proof employs the definition of information sets given in Gourieroux and Jasiak (2016), (2017). We derive the predictive density by using the Bayes theorem to condition the variables on their past and employ Jacobian formulas on manifolds.

1) Let us consider the information set:
$$I_{T+1} = (Y_1,...,Y_T, Y_{T+1}).$$
This set is equivalent to the set generated by $(Z_1, Z_2,...,Z_{T+1})$ and the set generated by $(Z_{1,2}, \eta_3, \eta_4,....,\eta_T, \eta_{1, T+1}, Z_{2, T}, Z_{2, T+1})$ by using the recursive equations (2.5). Since
$(  \eta_{1, T+1}, Z_{2, T}, Z_{2, T+1})$ is independent of $\underline{\varepsilon}_T$, we see that the conditional density $l(\eta_{1, T+1}, Z_{2, T}, Z_{2, T+1}|\underline{\varepsilon}_T) = l(\eta_{1, T+1}, Z_{2, T}, Z_{2, T+1})$ is equal to the marginal density.

It follows that the conditional density is:
\begin{equation}
l(\eta_{1, T+1}, Z_{2, T+1}|Z_{2, T}, \underline{\varepsilon}_T) = l(\eta_{1, T+1}, Z_{2, T+1}|I_T) =
l(\eta_{1, T+1}, Z_{2, T+1}|Z_{2,T}).
\end{equation}
\nin The last conditional density needs to be rewritten with a conditioning variable being the future $Z_2$. From the Bayes theorem, it follows that:
\begin{equation}
l(\eta_{1, T+1}, Z_{2, T+1}|I_T) = \frac{l_2(Z_{2, T+1})}{l_2(Z_{2, T})} \; l(\eta_{1, T+1}, Z_{2, T}|Z_{2,T+1}),
\end{equation}
\nin where $l_2$ is the marginal density of $Z_{2,t}$.

\medskip
2) Let us now consider the vector $\eta_t = A^{-1} \left(\begin{array}{c} \varepsilon_t \\ 0 \end{array}\right)$. This random vector takes values on the subspace $E = A^{-1} (\eR^m \times 0^{n-m})$. Its distribution  admits a density
$g_\eta(\eta_1, \eta_2)$ with respect to the Lebesgue measure on subspace $E$. Moreover, we have:
\begin{equation}
\eta_{T+1} = \left(\begin{array}{c} \eta_{1, T+1} \\ Z_{2, T+1} - J_2 Z_{2,T} \end{array}\right) =
\left(\begin{array}{cc} Id & 0 \\ 0 & -J_2 \end{array}\right) \left(\begin{array}{c} \eta_{1,T+1} \\ Z_{2,T} \end{array}\right) + \left(\begin{array}{c} 0 \\ Z_{2, T+1} \end{array}\right).
\end{equation}
\nin Then, conditional on $Z_{2, T+1}$, vector $\left(\begin{array}{c} \eta_{1,T+1} \\ Z_{2,T} \end{array}\right) $
takes values in the affine subspace \linebreak $F = \left(\begin{array}{cc} Id & 0 \\ 0 & -J_2 \end{array}\right)^{-1}
\left[ E - \left(\begin{array}{c}  0 \\  Z_{2, T+1} \end{array}\right) \right]$ with a density with respect to the Lebesgue measure on $F$. Since the transformation from $\eta_{T+1}$ to $\left(\begin{array}{c} \eta_{1,T+1} \\ Z_{2,T} \end{array}\right)$ is linear affine invertible, we can apply the Jacobian formula to get:
\begin{equation}
l(\eta_{1, T+1}, Z_{2, T}|Z_{2, T+1}) = |\det J_2|\, g_{\eta} (\eta_{1, T+1}, Z_{2, T+1} - J_2 Z_{2, T}).
\end{equation}
\nin Then from (a.2), (a.4) and $Z_{1, T+1} = J_1 Z_{1, T} + \eta_{1, T+1}$, it follows that:
\begin{equation}
l(Z_{1, T+1}, Z_{2, T+1}|I_T) = \frac{l_2(Z_{2, T+1})}{l_2(Z_{2, T})} \; |\det J_2|\, g_{\eta}( Z_{1, T+1}- J_1
Z_{1,T}, Z_{2,T+1} - J_2 Z_{2,T}).
\end{equation}

\nin Let us now derive the predictive density of $Y_{T+1}$ given $I_T$. We get a succession of affine transformations of variables with values in different affine subspaces (depending on the conditioning set) along the following scheme:
$$ \begin{array}{ccccccc}
\left(\begin{array}{c} \varepsilon_{T+1} \\ 0 \end{array}\right) & \stackrel{A^{-1}}{\rightarrow} & \eta_{T+1} &
\stackrel{Id}{\rightarrow} & Z_{T+1} & \stackrel{A}{\rightarrow} & \left(\begin{array}{c} Y_{T+1} \\ \tilde{Y}_{T} \end{array}\right).\\
 & & & &  (\mbox{given} \; Z_T) & & (\mbox{given} \; \tilde{Y}_{T})
\end{array}
$$
\nin Then, we can apply three times the Jacobian formula on manifolds. Since $|\det A^{-1}| |\det A| = \frac{|\det A|}{|\det A|} =1$,
the Jacobians cancel out and the predictive density becomes:
$$ l(y| \underline{Y}_T ) = \frac{l_2 \left[A^2 \left(\begin{array}{c} y \\ \tilde{Y}_{T} \end{array}\right)\right]}{l_2 \left[ A^2
\left(\begin{array}{c} Y_{T} \\ \tilde{Y}_{T-1} \end{array}\right) \right]} \, |\det J_2 | \, g(y- \Phi_1 Y_T - \cdots - \Phi_p Y_{T-p+1}),$$
\nin which yields the formula in Proposition 1.

In addition, from (a.5) we  derive the predictive density of $Z_{T+1}$ given $\underline{Z}_T$ as:
 \begin{equation}
l(Z_{T+1}|\underline{Z}_T) = \frac{l_2 (Z_{2,T+1})}{l_2 (Z_{2,T})} \; |det J_2| \; |det A| \;\;
g \left( A \left[ \begin{array}{c}  Z_{1,T+1} - J_1 Z_{1,T} \\ Z_{2,T+1} - J_2 Z_{2,T} 
\end{array}    \right] \right).
 \end{equation}
The predictive density of $Z_{T+1}$ depends on the choice of the state space representation, whereas the predictive density of $Y_T$ does not.

\begin{center}
{\bf Proof of Corollary 2}
\end{center}

To keep the notation simple, let us assume a mixed VAR(1) model. Then, from Corollary 1, it follows that $(Y_t)$ as well as $(Z_t)$ are Markov processes of order 1 in both calendar and reverse time. The distribution of process $(Z_t)$ is characterized by the pairwise distribution of $(Z_{t-1}, Z_t)$.

From the proof of Proposition 1, it follows that this joint distribution is:
$$
l(z_{t-1}, z_t) = l_1(z_{1, t-1}) l_2(z_{2,t})\; |det\, J_2| \; g_{\eta}(z_{1,t} - J_1 z_{1, t-1}, 
z_{2,t} - J_2 z_{2, t-1}).$$
\nin Then, the conditional distribution of $Z_{t-1}$ given $Z_t=z_t$  is:
\begin{eqnarray*}
l(z_{t-1}|z_t) & = & l(z_{t-1}, z_t)/l(z_t) \\
& = & l(z_{t-1}, z_t) / [ l_1(z_{1,t}) \, l_2(z_{2,t})] \;\mbox{,because} \; Z_{1,t} \; \mbox{and} \;
Z_{2,t} \; \mbox{are independent,} \\
& = & \frac{l_1(z_{1,t-1})}{l_1(z_{1,t})} \; |det \; J_2| \; g_{\eta} (z_{1,t} - J_1 z_{1, t-1}, 
z_{2,t} - J_2 z_{2, t-1}).
\end{eqnarray*}
\nin The result in Corollary 2 follows by applying the transformations: $Y_t = AZ_t, \; \varepsilon_t = A \eta_t$.

\begin{center}

APPENDIX A.2 \\

{\bf The Multiplicative Causal-Noncausal Model}
\end{center}

A constrained  multiplicative Mixed Autoregressive (MAR) representation of the mixed VAR model was proposed by Lanne, Saikkonen (2013):
\begin{equation}
\Phi(L) \Psi (L^{-1}) Y_t = \epsilon_t^*,
\end{equation}
\nin where the matrices of autoregressive polynomials $\Phi$ and $\Psi$ have both roots outside the unit circle, such that $\Psi(0) = \Phi(0) = Id$ and $\epsilon_t^*$
is an i.i.d. sequence of errors. The multiplicative representation exists for the univariate mixed processes [Breidt et al. (1991), Lanne and Saikkonen (2008)]. For multivariate processes, Davis and Song (2020),  Swensen (2022), point out that the multiplicative representation of a mixed causal-noncausal VAR model does not always exist and  implies restrictions on the coefficient matrices $\Phi_1,...,\Phi_p$.

%Otherwise, Online Appendix A.2 shows that the multiplicative representation may not be compatible with the representation (2.1) of the
%VAR model. For example, the multiplicative VAR(2) model $(Id-\Phi L) (Id - \Psi L^{-1}) Y_t = \varepsilon_t^*$ where the matrix polynomials have roots of modulus strictly greater than 1 and $(\varepsilon_t^*)$ are i.i.d., can be written under the form $Y_t = \Phi_1 Y_{t-1} + \Phi_2 Y_{t-2} + \varepsilon_t$. However, the errors $\varepsilon_t$ are independent only if the matrix $\Psi $ is invertible.

In fact, the set of multiplicative VAR and mixed VAR models are non-nested, i.e. a mixed model does not always admit a multiplicative form and a multiplicative VAR model cannot always be written as a mixed VAR.

This major difficulty is a consequence of a different normalization. For example, if $\Phi(L) = Id - \Phi L$
and $\Psi(L^{-1}) = Id - \Psi L^{-1}$, then the multiplicative model is such that:
$$ \Phi(L) \Psi(L^{-1}) Y_t = -\Psi Y_{t+1} + (Id + \Phi \Psi) Y_t  - \Phi Y_{t-1} = \varepsilon_t^*,$$
\nin which cannot be transformed into:
$$Y_t = \Phi_1 Y_{t-1} + \Phi_2 Y_{t-2} + \varepsilon_t,$$
\nin with i.i.d. errors, if matrix $\Psi$ is not invertible.
Let us now consider the reciprocal problem. We have the following proposition:

\nin {\bf Proposition A.1:} When $m \geq 2$, a mixed VAR(1) representation with at least one causal and one non-causal roots cannot be factorized as in (a.7).

{\bf Proof:} For ease of exposition, we consider below the bivariate case $m=2$. The results remain valid for $m \geq 2$. Let us consider a mixed VAR(1) model
$$Y_t = \Phi_1 Y_{t-1} + \epsilon_t,$$

\nin where $\Phi_1$ has eigenvalues $\lambda_1<1$ and $\lambda_2>1$. If it can be written in a multiplicative form (a.7), we deduce from the stationarity condition of the process that in this decomposition $det(Id - \Phi z)$ [resp. $det(Id - \Psi z)$] has a unique root $\lambda_1$ (resp. $1/\lambda_2$).

\nin However, we have: $det(Id - \Phi z) = 1-z Tr \Phi + z^2 det \Phi,$
and
$det(Id - \Psi z) = 1-z Tr \Psi + z^2 det \Psi.$
\nin These polynomials have a unique root if and only of $det \Phi = det \Psi =0$, that is if $\Phi$ and $\Psi$ are non-zero and their rank is equal to 1. Then, we have:
$$-\Psi Y_t +(Id + \Phi \Psi)Y_{t-1} - \Phi Y_{t-2} = \epsilon_t^*$$
\nin  We cannot normalize this equation as in (2.1) since $\Psi$ is not of full rank, and we cannot obtain a VAR(1) model since $\Phi \neq 0$.

\newpage

\begin{center}
{\bf Online APPENDICES \\

B: Predictive Density Estimation}
\end{center}

This Appendix describes the kernel-based estimation of the predictive density given in Proposition 1 from the following time series:

$$\hat{\varepsilon}_t = Y_t - \hat{\Phi}_1 Y_{t-1} - \cdots - \hat{\Phi}_p Y_{t-p+1}, \;\; t=1,...,T,$$
$$\hat{Z}_{2,t} = \hat{A}^2 \left( \begin{array}{c} Y_t  \\ \tilde{Y}_{t-1}  \end{array} \right), \;\; t=1,...,T.$$

\nin The above time series are used to approximate the density $g$ of $\varepsilon_t$ and density $l_2$ of
$Z_{2,t}$ as follows:

$$\hat{g}_T (\varepsilon) = \frac{1}{T} \frac{1}{h_1^m} \sum_{t=1}^T K_m \left( \frac{\varepsilon- \hat{\varepsilon}_t}{h_1}   \right),$$

\nin and

$$ \hat{l}_{2,T} (z_2)  = \frac{1}{T} \frac{1}{h_2^{n_2}} \sum_{t=1}^T K_{n_2} \left( \frac{z_2- \hat{Z}_{2,t}}{h_2}   \right),$$

\nin where $h_1, h_2$ are the bandwidths and $K_m, K_{n_2}$ are multivariate kernels of dimensions $m$ and $n_2$, respectively. Then, the estimated predictive density is:

$$ \hat{l}_T (y| \underline{Y}_T ) = \frac{\hat{l}_{2,T} \left[\hat{A}^2 \left(\begin{array}{c} y \\ \tilde{Y}_{T} \end{array}\right)\right]}{\hat{l}_{2,T} \left[ \hat{A}^2
\left(\begin{array}{c} Y_{T} \\ \tilde{Y}_{T-1} \end{array}\right) \right]} \, |\det \hat{J}_2 | \, \hat{g}_T (y- \hat{\Phi}_1 Y_T - \cdots - \hat{\Phi}_p Y_{T-p+1}),$$

This formula is easily extended to bandwidths adjusted for each component by replacing for example
$ \frac{1}{h_1^m} K_m \left( \frac{\varepsilon- \hat{\varepsilon}_t}{h_1}   \right)$ by $\prod_{j=1}^m
\frac{1}{h_{1j}} K \left( \frac{\varepsilon_j- \hat{\varepsilon}_{j,t}}{h_{1j}}   \right)$,
where $K$ is a univariate kernel. Such an adjustment can account for different component variances.

Let us consider the example of a bivariate VAR(1) process with one noncausal component and a scalar  noncausal eigenvalue $\lambda_2$ (see Section 6). 
The estimated coefficients of the inverse of $A$ are denoted by:

$$\hat{A}^{-1} = \left( \begin{array}{cc} \hat{a}^{11} & \hat{a}^{12}  \\
\hat{a}^{21} & \hat{a}^{22}  \end{array} \right).$$

The predictive density depends on unknown scalar parameters $\lambda_1, \lambda_2$ and functional parameters $l_2, g$ that can be estimated. The marginal density 
$l_2 (A^2 y)$
can be approximated by 
a kernel estimator:

$$\hat{l}_{2,T} (\hat{A}^2 y) = \frac{1}{T} \frac{1}{h_2} \sum_{t=1}^T K \left( \frac{\hat{a}^{21}(y_1 - y_{1,t}) + \hat{a}^{22}(y_2 - y_{2,t})}{h_2} \right),$$

\nin while the density $l_2 ( A^2 Y_{T})$
can be approximated by 
a kernel estimator:

$$\hat{l}_{2,T} (y_T) = \frac{1}{T} \frac{1}{h_2} \sum_{t=1}^T K \left( \frac{\hat{a}^{21}(y_{1,T} - y_{1,t}) + \hat{a}^{22}(y_{2,T} - y_{2,t})}{h_2} \right),$$

\noindent where $h_2$ is a bandwidth. The joint density $g(y - \Phi y_T)$ can be approximated by

$$\hat{g}_T (y - \hat{\Phi} y_T)  = \frac{1}{T} \frac{1}{h_{11}h_{12}} \sum_{t=1}^T
K \left( \frac{y_1 - \hat{\phi}_{1,1} y_{1,T} - \hat{\phi}_{1,2} y_{2,T}  - \hat{\varepsilon}_{1,t} }{h_{11}} \right)
K \left( \frac{ y_2 - \phi_{2,1} y_{1,T} - \phi_{2,2} y_{2,T}  - \hat{\varepsilon}_{2,t} }{h_{12}} \right).$$

\noindent where $\hat{\varepsilon}_{1,t}$ and $\hat{\varepsilon}_{2,t}$ are residuals
$\hat{\varepsilon}_t = y_t - \hat{\Phi} y_{t-1}$
 and $h_{11},h_{12}$ are two bandwidths adjusted for the variations of $\hat{\varepsilon}_{1,t}$ and $\hat{\varepsilon}_{2,t}$, respectively. We get:

\begin{eqnarray*}
\hat{l}_T (y_1, y_2|Y_T) & = & \frac{\frac{1}{T} \frac{1}{h_2} \sum_{t=1}^T K \left( \frac{\hat{a}^{21}(y_1 - y_{1,t}) + \hat{a}^{22}(y_2 - y_{2,t})}{h_2} \right)}{
\frac{1}{T} \frac{1}{h_2} \sum_{t=1}^T K \left( \frac{\hat{a}^{21}(y_{1,T} - y_{1,t}) + \hat{a}^{22}(y_{2,T} - y_{2,t}) }{h_2} \right)} \\
& & \times \; |\hat{\lambda}_2| \frac{1}{T} \frac{1}{h_{11} h_{12}} \sum_{t=1}^T
K \left( \frac{ y_1 - \hat{\phi}_{1,1} y_{1,T} - \hat{\phi}_{1,2} y_{2,T}  - \hat{\varepsilon}_{1,t}) }{h_{11}} \right) \\
& & \times \; K \left( \frac{ y_2 - \hat{\phi}_{2,1} y_{1,T} - \hat{\phi}_{2,2} y_{2,T}  - \hat{\varepsilon}_{2,t}) }{h_{12}} \right).
\end{eqnarray*}

\bigskip

\begin{center}

{\bf C: Nonparametric Estimation of Conditional Quantiles of Filtered Variables}
\end{center}

We derive the closed form expression of the estimator of the conditional quantile based on observations $Z_{2,t}, t=1,...,T$. When  $Z_{2,t}$ is not observed but filtered, the estimator is modified by replacing in the formula  $Z_{2,t}$ by  $\hat{Z}_{2,t}$.

i) Conditional cdf

The conditional cdf $F_2(x,z) = P[ Z_{2,t} \leq x|  Z_{2,t-1} = z]$ is estimated by the Nadaraya-Watson estimator:

$$
\hat{F}_2(x,z) = \sum_{t=2}^T \left[ 1_{Z_{2,t} \leq x} K \left( \frac{Z_{2,t} -z}{h} \right) \right]/\sum_{t=2}^T  K \left( \frac{Z_{2,t} -z}{h} \right),$$
 
 \nin where $K(.)$ denoted the kernel and $h$ the bandwidth.
 
 This estimated conditional cdf is a piecewise function, with values obtained as follows:
 
 step 1: rank the observations $Z_{2,t}, t=2,...,T$, as $Z_{2,(h)}, h=1,...,T-1$
 \footnote{We assume that these values are distinct. Otherwise, we make them distinct by adding a small positive constant.}.

 step 2: The values of $\hat{F}_2(x,z)$ are $p(h,z) = \hat{F}_2( Z_{2, (h)},z), h=1,...,T-1$. This function is strictly increasing in $h$.

ii) Conditional Quantile

We invert the conditional cdf and evaluate its inverse at probability $\alpha= \Phi(v_2)$. We get the following closed form estimator

$\hat{G}_2 (z, v_2) = Z_{2,k(z,v_2)}$ where:

$$
k(z,v_2)  = \left\{ \begin{array}{ll} 1, & \mbox{if} \; \; \Phi(v_2) < p(1,z). \\
inf \{ h: p(h,z) < \Phi(v_2)\}, & \mbox{if} \;\;  \Phi(v_2) > p(1,z).
\end{array} \right. $$
  
 \bigskip
\setcounter{figure}{0}\def\thefigure{a.\arabic{figure}}
\setcounter{table}{0}\def\thetable{a.\arabic{table}}

\begin{center}
{\bf D: Additional Simulation Results}
\end{center}

This Appendix presents additional results on predictions and prediction intervals based on the simulated bivariate process displayed and estimated in Section 6.1.
We consider an extend simulated path of 600 observations displayed in Figure a.1: 

\begin{center}
\begin{figure}[h]
\centering
\includegraphics[width=12cm,height =3cm]{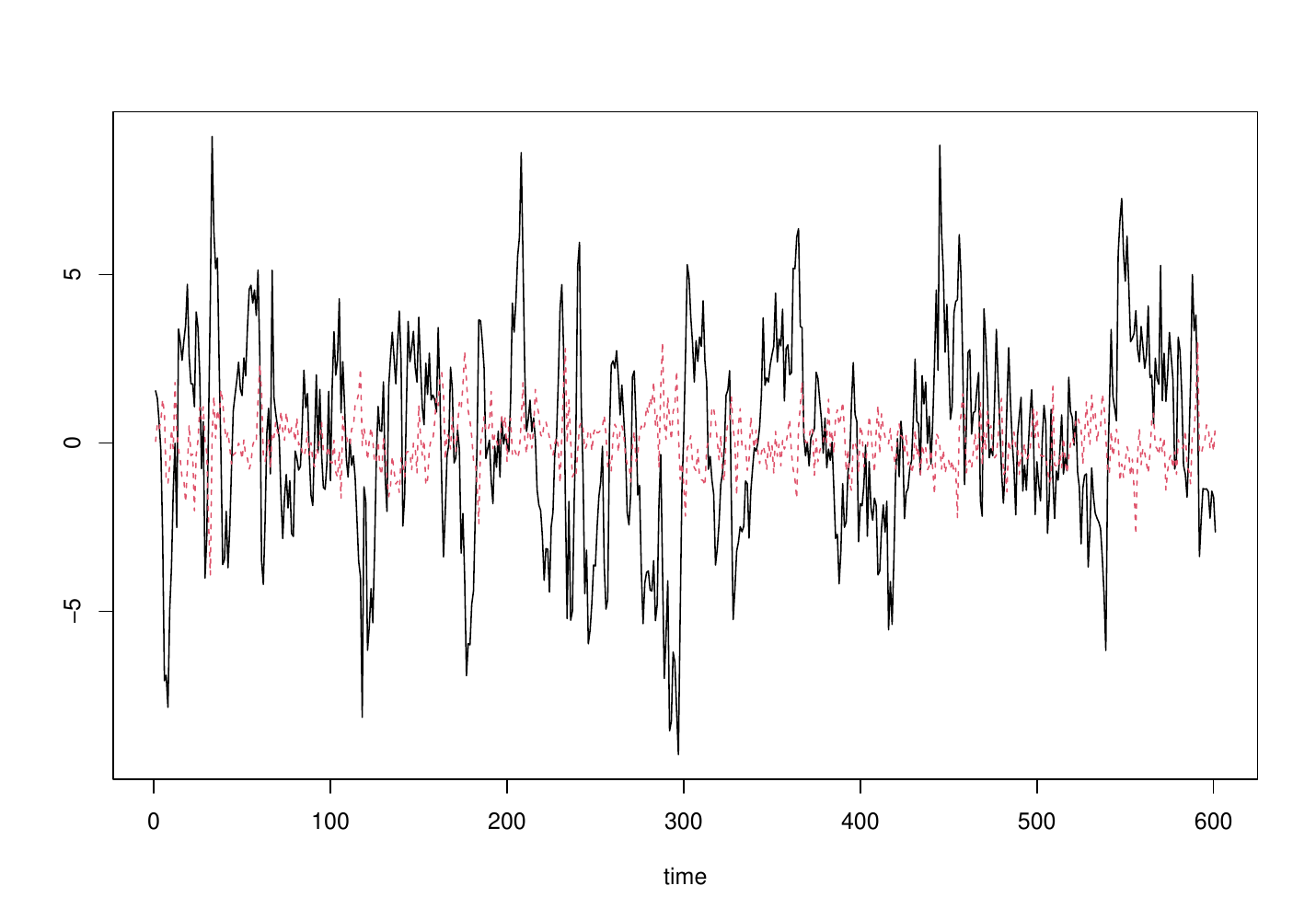}
\caption{Bivariate Mixed VAR(1) Process: $Y_1$: solid line, $Y_2$: dashed line}
\end{figure}
\end{center}

%\begin{center}
%\begin{figure}[h]
%\centering
%\includegraphics[width=12cm,height =3cm]{Figcompofromestimated.pdf}
%\caption{Bivariate Mixed VAR(1) Process: $Y_1$: solid line, $Y_2$: dashed line}
%\end{figure}
%\end{center}

\nin Next, we compute oos a series of one-step ahead forecasts starting from $Y_{500}$, based on the estimators obtained from the first 500 observations. The series of 100 one-step ahead forecasts along the trajectory is displayed in Figure a.2.

\begin{figure}
  \centering
  \begin{subfigure}[b]{\linewidth}
  \centering
\includegraphics[width=0.7\linewidth, height =3cm]{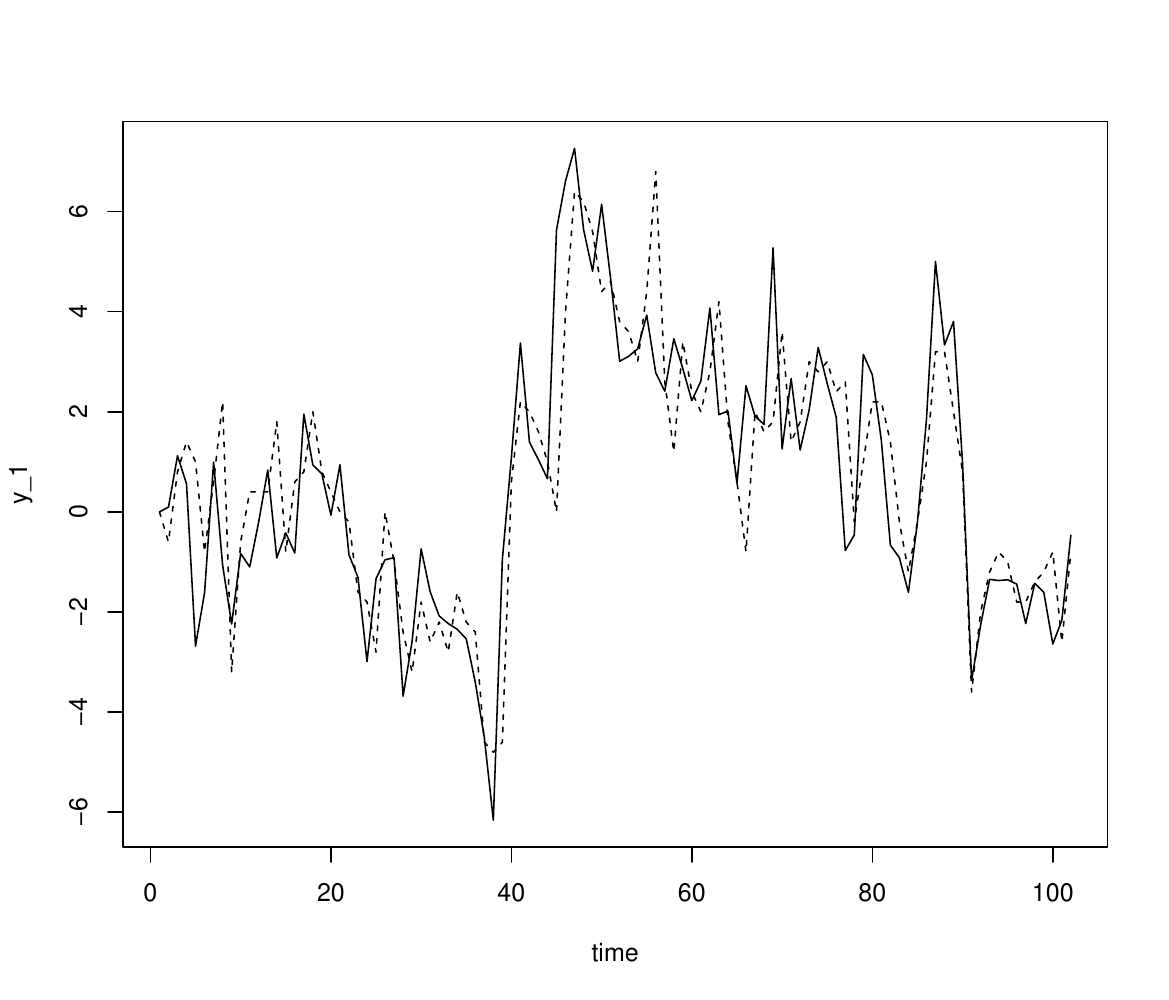}
     \caption{Estimated 1-step-ahead Forecasts of $Y_{1,t}$}
  \end{subfigure}
  \centering
  \begin{subfigure}[b]{\linewidth}
  \centering
\includegraphics[width=0.7\linewidth, height =3cm]{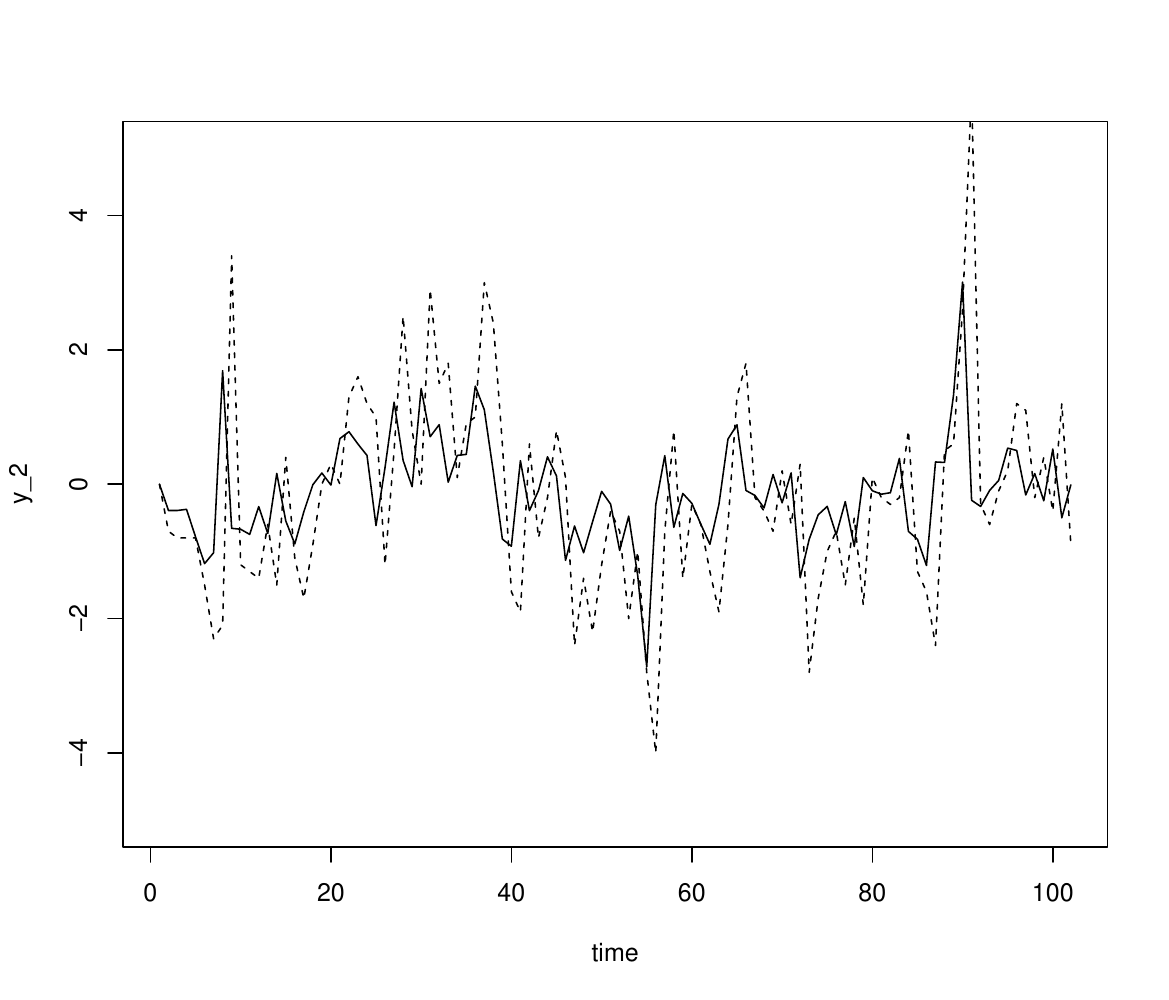}
    \caption{Estimated 1-step-ahead Forecasts of $Y_{2,t}$}
  \end{subfigure}
\caption{Estimation-Based oos One-Step Ahead Forecasts true:solid line, forecast:dashed line}
\end{figure}

We observe that the predictions are reliable even during the bubble episodes and on spikes. The predictions are also relatively better for series $Y_1$ with longer episodes of local trends.

\nin{\bf Unconditional Coverage of  the Estimated  Prediction Interval}

Let us now consider two data generating (DGP) processes. The first one is the bivariate mixed VAR(1) illustrated in 6.1 with the autoregressive matrix:
$\Phi = \left( \begin{array}{cc} 0.7 & -1.3 \\ 0.0 & 2.0 \end{array} \right),$
eigenvalues $\lambda_1 = 0.7$ and $\lambda_2=2.0$.
%and an identity variance-covariance matrix of errors $\varepsilon_t$. 
In this experiment, the errors have t-student distributions with 3, 6 and 9 degrees of freedom, respectively. The second process is a bivariate VAR(1) with the autoregressive matrix $\Phi = \left( \begin{array}{cc} 0.9 & -0.3 \\ 0.0 & 1.2 \end{array} \right),$
and eigenvalues $\lambda_1 = 0.9, \lambda_2= 1.2$, which are closer to the unit circle. The errors  $\varepsilon_t$ have 
%an identity variance-covariance matrices and 
t-student distributions with 3, 6 and 9 degrees of freedom.

The simulated bivariate series are of length 100, 500 and 1000 and each DGP is
replicated 500 times. The last observation from each simulated path is set aside. It is forecast and used for forecast coverage assessment.
The autoregressive parameters $\hat{\Phi}$ are estimated by the GCov estimator from the autocovariances of errors $\varepsilon_t = Y_t - \Phi Y_{t-1}$, and of their second, third and fourth powers, up to lag $H=10$. The algorithm maximizing the GCov objective function is initiated each time at the starting values 0.1, 0.1, 0.5, 0.5.
The oos predictive density of $y(T+1)$ given in (3.2) is evaluated from the parameter estimates, over a grid of 100 possible future values for each of the components series. We use Gaussian kernels and bandwidths $h_2=s.d.(Y_{2,t})$
$h_{11} = s.d.(\varepsilon_1)$ and $h_{12} = s.d.(\varepsilon_2)$ (see Online Appendix B for kernel density estimators). The main mode of the predictive density, estimated from formula (3.2) with Gaussian kernels, provides the point forecast. 
The oos prediction intervals at horizon 1 are obtained from the marginal 10th and 90th percentiles of the bivariate predictive density. As before, the rationale for choosing level 80\% is to ensure a sufficiently large number of observations in the tails for computing the quantiles of predictive densities.

The unconditional coverage of the estimated prediction interval is reported in Table a.1 below for the two DGPs described above, sample sizes T=500 and T=1000
and t-student distributions of error $\varepsilon_t$ with 3, 6 and 9 degrees of freedom. We observe that it is either

%\singlespacing
\begin{table}

\begin{center}
\begin{tabular}{|l|c|c|c||c|c|c||c|c|c|}
\hline
\multicolumn{1}{|c|}{}&\multicolumn{3}{|c|}{T=100}&\multicolumn{3}{|c|}{T=500}&\multicolumn{3}{|c|}{T=1000} \\
\hline
\multicolumn{10}{|c|}{VAR(1) with eigenvalues 0.7, 2.0}\\
\hline
component & t(3) & t(6) & t(9) & t(3) & t(6) & t(9) &  t(3) & t(6) & t(9) \\ \hline
$y_1(T+1)$ & 90.0 & 91.0 & 92.6 & 91.4 & 92.8 & 92.0 & 91.4 & 90.8 & 93.2  \\
$y_2(T+1)$ & 92.2 & 93.8 &  91.8 & 93.4 & 94.0 & 94.8 & 95.2 &  93.2 & 94.5  \\ \hline
\multicolumn{10}{|c|}{VAR(1) with eigenvalues 0.9, 1.2}\\
\hline
component & t(3) & t(6) & t(9) & t(3) & t(6) & t(9) &  t(3) & t(6) & t(9)\\ \hline
$y_1(T+1)$ &  80.2  & 82.8& 87.4 &   82.2 & 86.0 & 84.2 & 84.6  & 84.4 & 88.6 \\
$y_2(T+1)$ &  90.4 &89.4 & 90.8  &  90.6 & 91.6 & 92.40 & 91.4 & 91.0  & 92.8 \\
\hline
\end{tabular}
\caption{Unconditional Coverage of GCov Estimated Prediction Interval at 80\% \\
Based on the sample quantiles of estimated predictive densities from sample sizes of 100, 500 and 1000 observations (col.2:4) for t-student distributed errors with 3,6 and 9 degrees of freedom}

\end{center}

\end{table}

%\doublespacing

\nin   greater or close to the theoretical conditional coverage of the prediction
interval for both DGPs and sample sizes. It does not mean that the prediction interval is "conservative". The reported coverage rate depends not only on the prediction quality but also on the accuracy of the quantile estimator. Because the predictive density is approximated at a relatively low number of points (10000), has long tails and the tail observations are not dense, we use a robust quantile estimator, which approximates the quantiles "from above", resulting in practice in a unconditional coverage higher than the nominal conditional coverage rate. 

\medskip

\nin {\bf Estimated Prediction Set Uncertainty}

This Section illustrates the uncertainty in the estimated conditional prediction interval described
in Section 4.
We consider the simulated trajectory of length T=200 of the bivariate VAR(1) process illustrated in Table 1 with autoregressive matrix:  
$\Phi = \left( \begin{array}{cc} 0.9 & -0.3 \\ 0.0 & 1.2 \end{array} \right),$
and eigenvalues $\lambda_1 = 0.9, \lambda_2= 1.2$ and t(6) distributed errors with an identity variance-covariance matrix.
We are interested in the conditional prediction interval out-of-sample of the first component $Y_{1, T+1}$ = $Y_{1,200}$ given the past and current values of the two series.

The mixed VAR(1) model is estimated by the GCov estimator from the observations $t=1, ...,199$ with four power transforms of model errors and lag $H=10$, providing the following estimates of autoregressive parameters:
0.8931, -0.2146, 0.0180, 1.2797, and eigenvalues: 0.903 and 1.269. The estimated predictive density of $Y_{200}$ conditional on $Y_{199}$ is computed from formula (3.2) by using kernel smoothed density estimators. More specifically, we employ Gaussian kernels and bandwidths $h_2 = s.d.(Y_2), h_{11} = s.d. (\varepsilon_1), h_{12}=s.d(\varepsilon_2)$ (see Online Appendix B for the kernel density estimators). For $\alpha_1=0.2$ and $\Phi^{-1}(\alpha_1/2) = -1.28$, the estimated prediction interval of $Y_{1,T+1} = Y_{1,200}$ is:
$\widehat{PI}(y, \alpha_1) = [ -6.954, 0.360]$. It contains the true value of $Y_{1,200} = -0.696$.

In the next step, we replicate the initial paths $S=50$ times by backcasting from the bivariate
terminal condition $Y_{199}= [-1.188, 0.473]$. We use the backcasting formula given in Corollary 2, evaluated from estimated model parameters and kernel density estimators. We employ Gausian kernels and bandwidths $h_1 = s.d.(Y_{1}), h_{11} = s.d. (\varepsilon_1)$ and $h_{12} = s.d.(\varepsilon_2)$. 

The randomness of the backcast paths conditional on $Y_T$ is generated as follows: i) We first draw 100 values as if the components $Y_{1, T-1}, Y_{2, T-1}$ were independent conditional on $Y_T$. This is done by inverting the estimated conditional c.d.f.s of  $Y_{1, T-1}$ and $Y_{2, T-1}$ given $Y_T$. This provides the sampling with the importance (misspecified) density; ii) Next, we re-sample in the set of values obtained in step i) above, with the weights proportional to the ratio of the joint backward predictive density divided by the product of the two marginal backward predictive densities. This procedure adjusts for the omitted cross-sectional dependence in step i). 

The parameter $\Phi$ and functional parameter $g$ are re-estimated from each replicated path. Next,
the values of $\hat{m}^s$, $\hat{\sigma}^s$ are computed from the quantiles of $50$ predictive densities of $Y_{T+1}$ conditional on $Y_T$.
Then, we obtain the bootstrap confidence interval at level $1-\alpha_2 = 0.95$ of the prediction interval $\widehat{CSPI}(y_1, \alpha_1, \alpha_2) = [ -11.839, 5.246],$ with the solution $\hat{q}(y, \alpha_1,\alpha_2)= 2.99$. The confidence interval 
$\widehat{CSPI}$ of the prediction interval  is much larger than the interval $\widehat{PI}$. This result illustrates the possibility of extending the level of the prediction interval from 80\% to 95\% while taking into account the estimation risk on $\Phi$ and $g$ given that the estimator of the functional parameter $g$ converges at a lower speed than the matrix parameter estimator. The extended interval is such that
the length of the interval has almost doubled and the interval became less symmetric with respect to 0.

\bigskip

\setcounter{equation}{0}\def\theequation{b.\arabic{equation}}
\begin{center}
{\bf E: Additional Empirical Results}
\end{center}

\nin {\bf 1. Oil Prices and US GDP Rates Estimation}

\nin The quarterly macroeconomic time series are displayed in Figure a.3.

\begin{figure}[h]
\centering
\includegraphics[width=12cm,height =5cm, angle = 0]{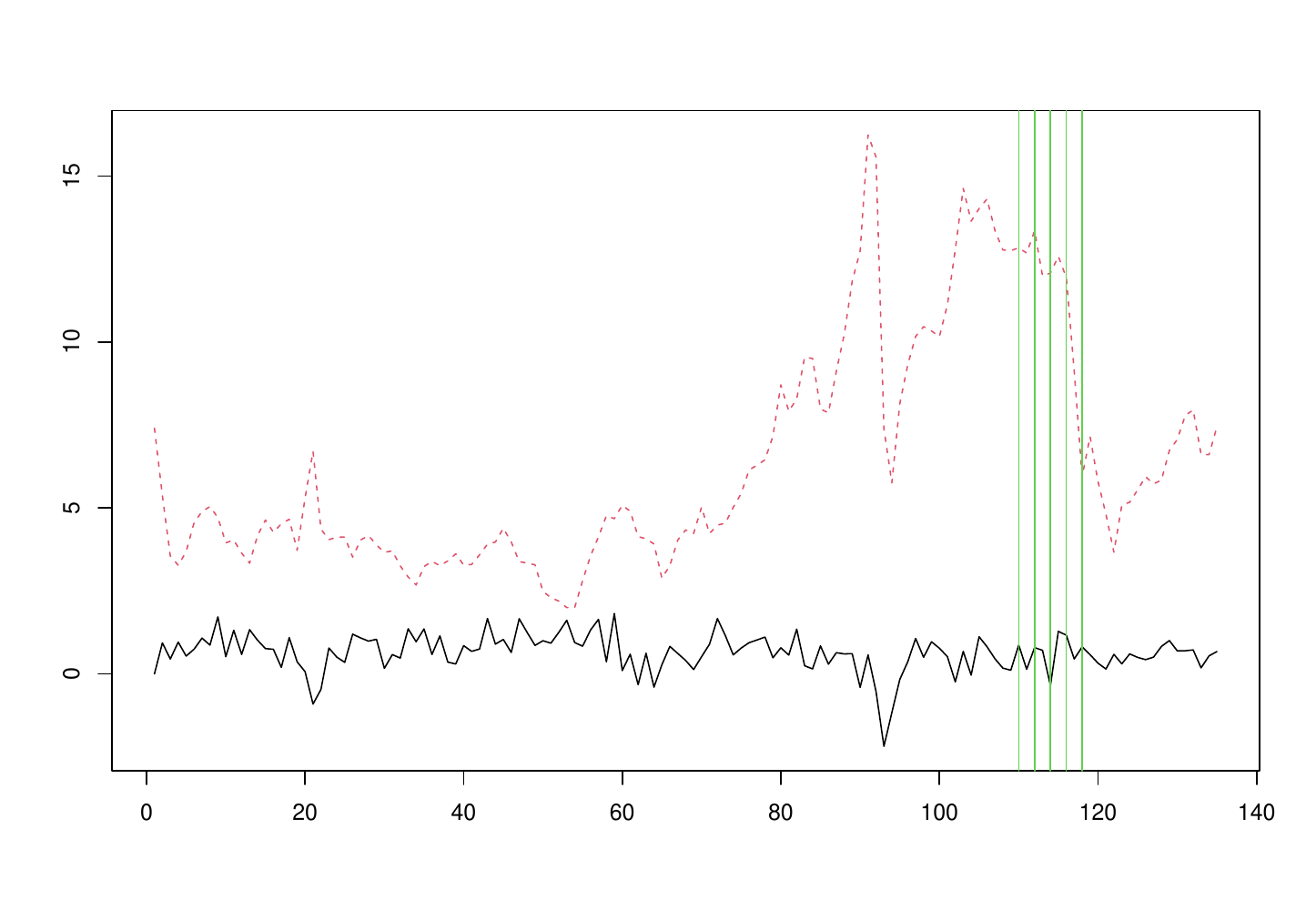}
\caption{The path of Oil and US GDP rate with prediction dates }
\end{figure}

\nin The green vertical lines indicate the prediction points discussed in the main text.

\nin {\bf 1.1 Mixed VAR(1): Estimation results} 
 
\nin The estimated errors of the mixed VAR(1) model are plotted in Figure a.4.

\begin{figure}[h]
\centering
\includegraphics[width=12cm,height =7cm, angle = 0]{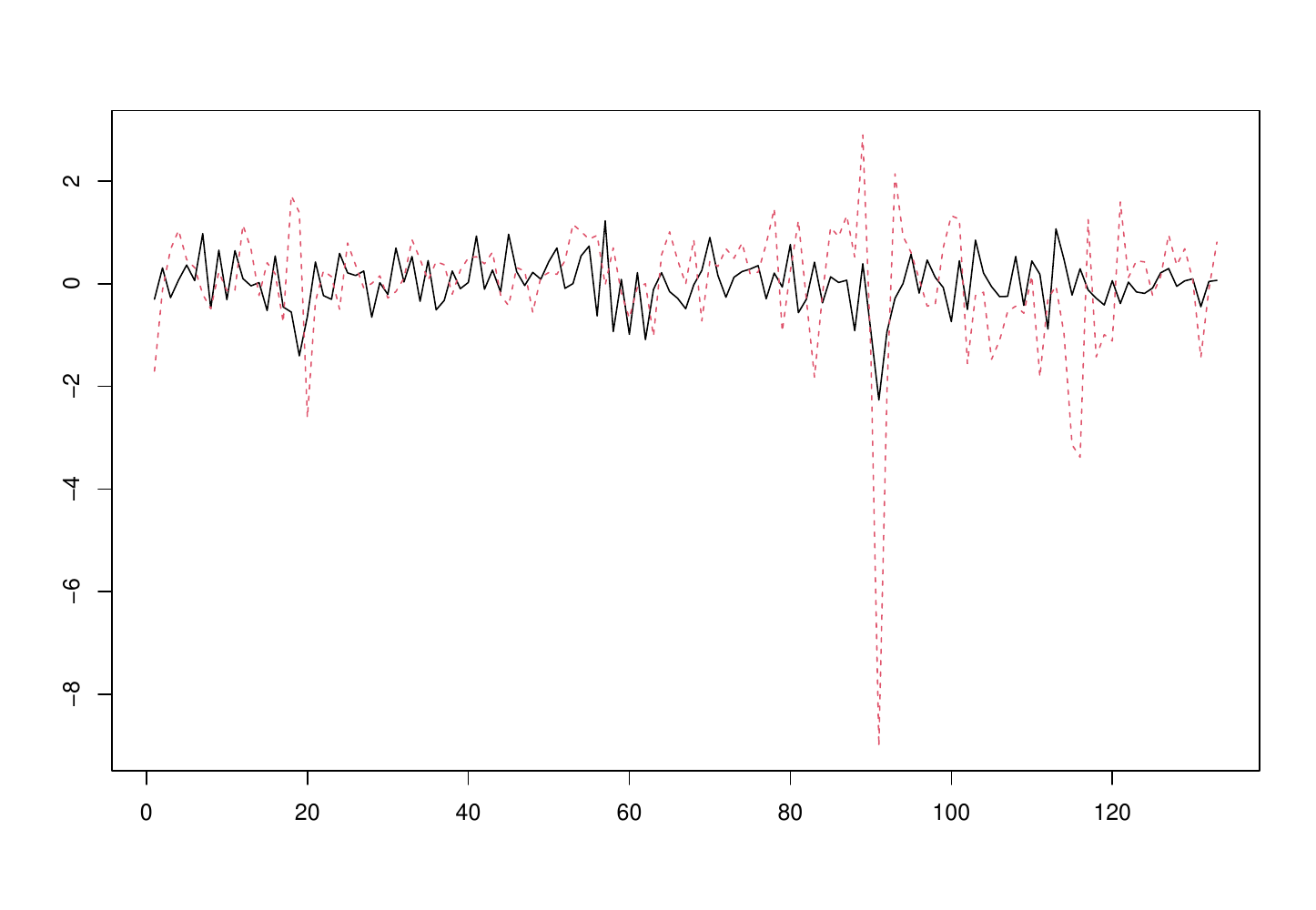}
\caption{Errors $\hat{\epsilon}_t$, Noncausal VAR(1), Q1 1986 -Q2 2019}
\end{figure}

\nin We observe that their densities are non-Gaussian, as shown in Figure a.5:

%\clearpage
%\newpage
\begin{figure}[h]
  \centering
  \begin{subfigure}[b]{0.4\linewidth}
\includegraphics[width=\linewidth]{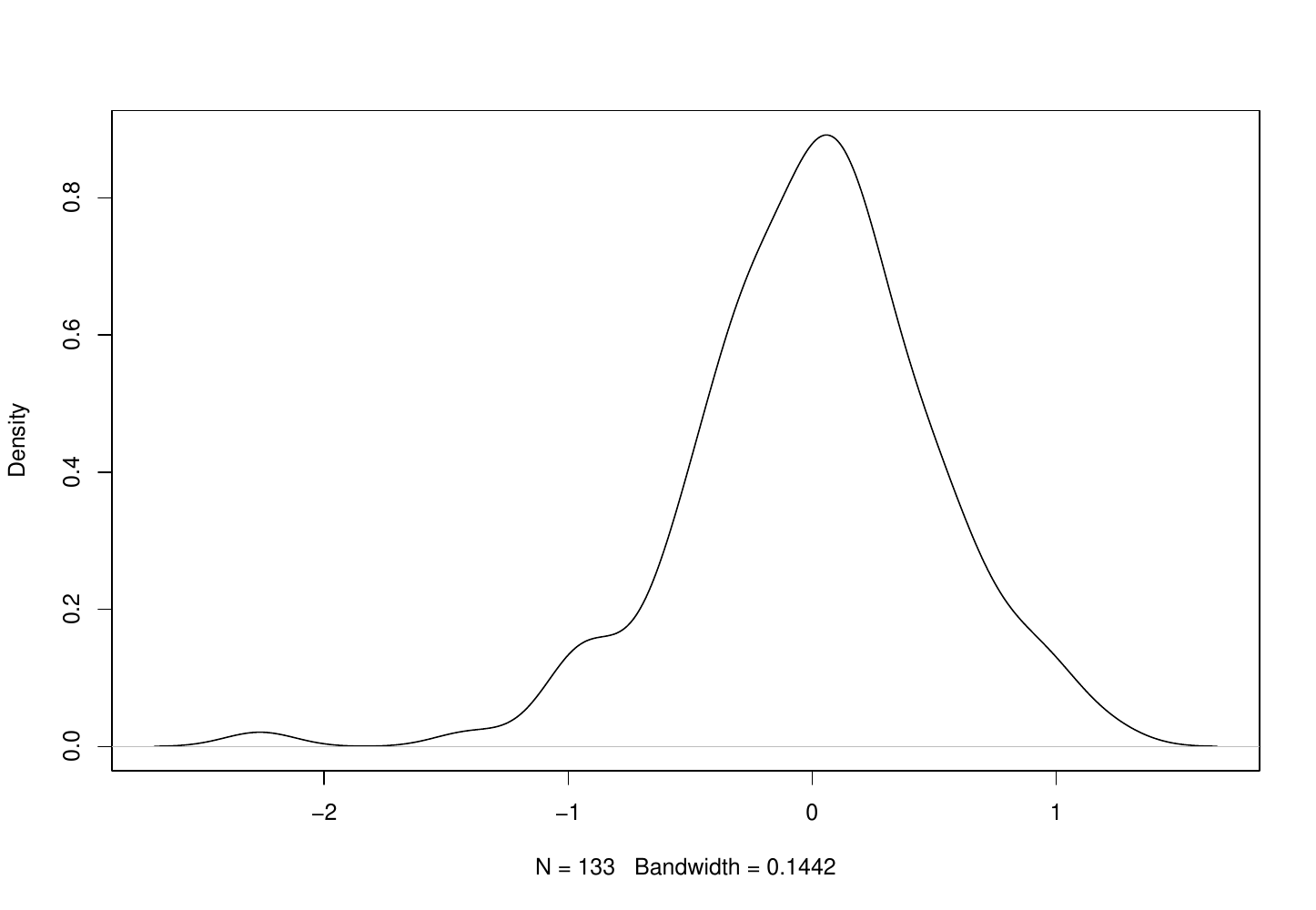}
     \caption{Density of $\hat{\epsilon}_{1,t}$}
  \end{subfigure}
  \begin{subfigure}[b]{0.4\linewidth}
\includegraphics[width=\linewidth]{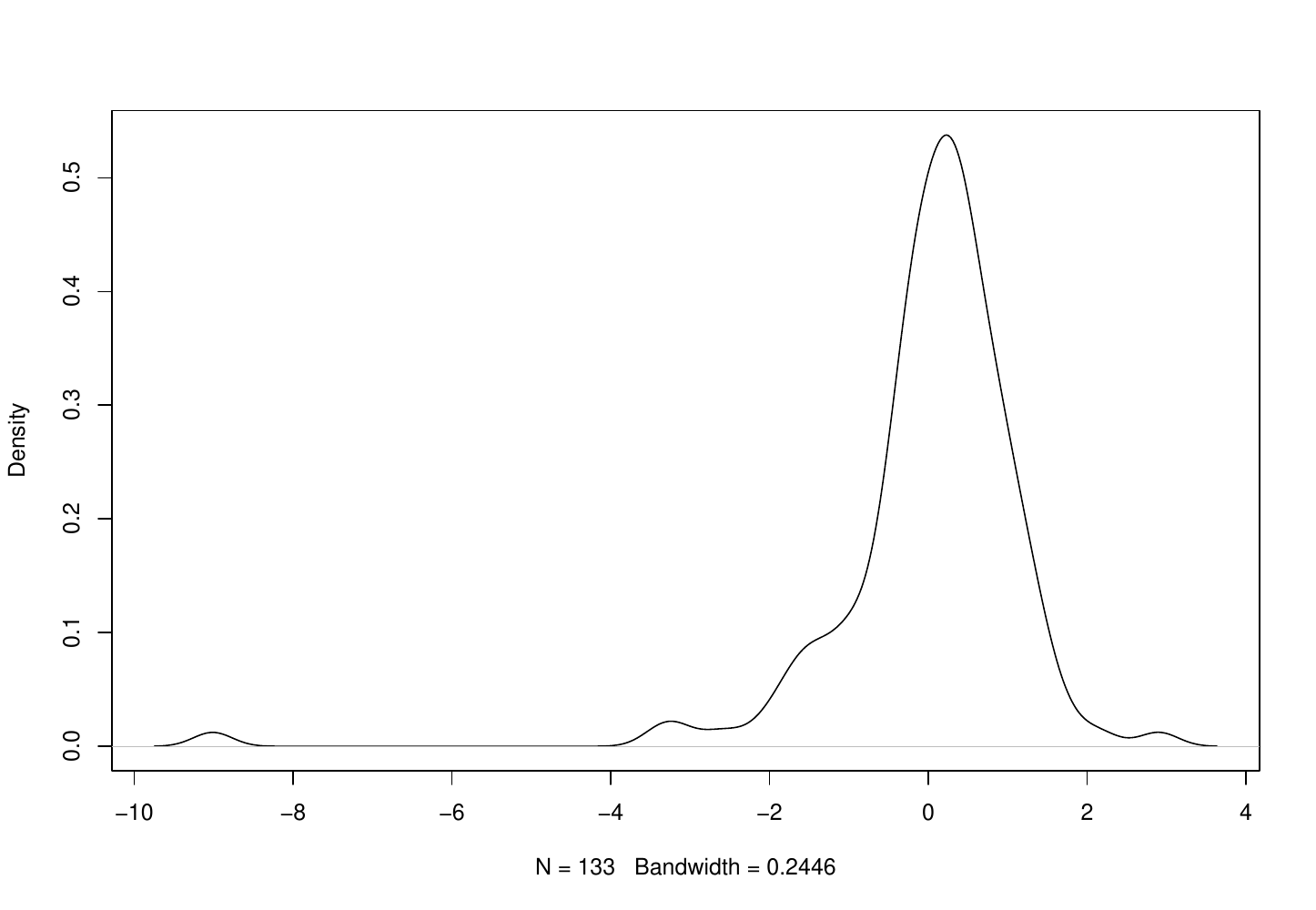}
    \caption{Density of $\hat{\epsilon}_{2,t}$}
  \end{subfigure}
\caption{Density of $\hat{\epsilon}_t$}
\end{figure}

\clearpage
%\newpage
\begin{figure}[h]
  \centering
  \begin{subfigure}[b]{0.4\linewidth}
\includegraphics[width=\linewidth]{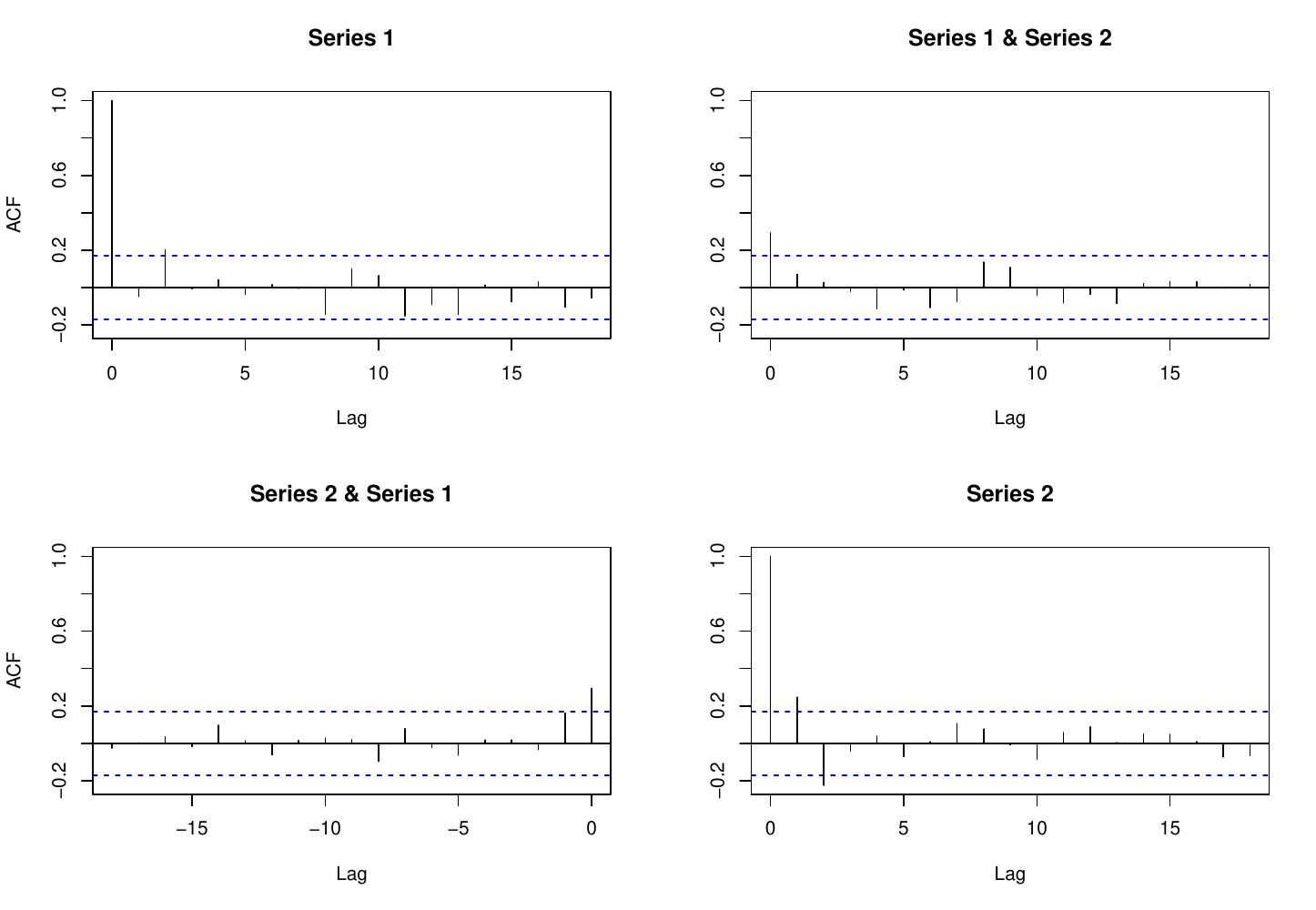}
     \caption{ACF of $\hat{\varepsilon}$}
  \end{subfigure}
  \begin{subfigure}[b]{0.4\linewidth}
\includegraphics[width=\linewidth]{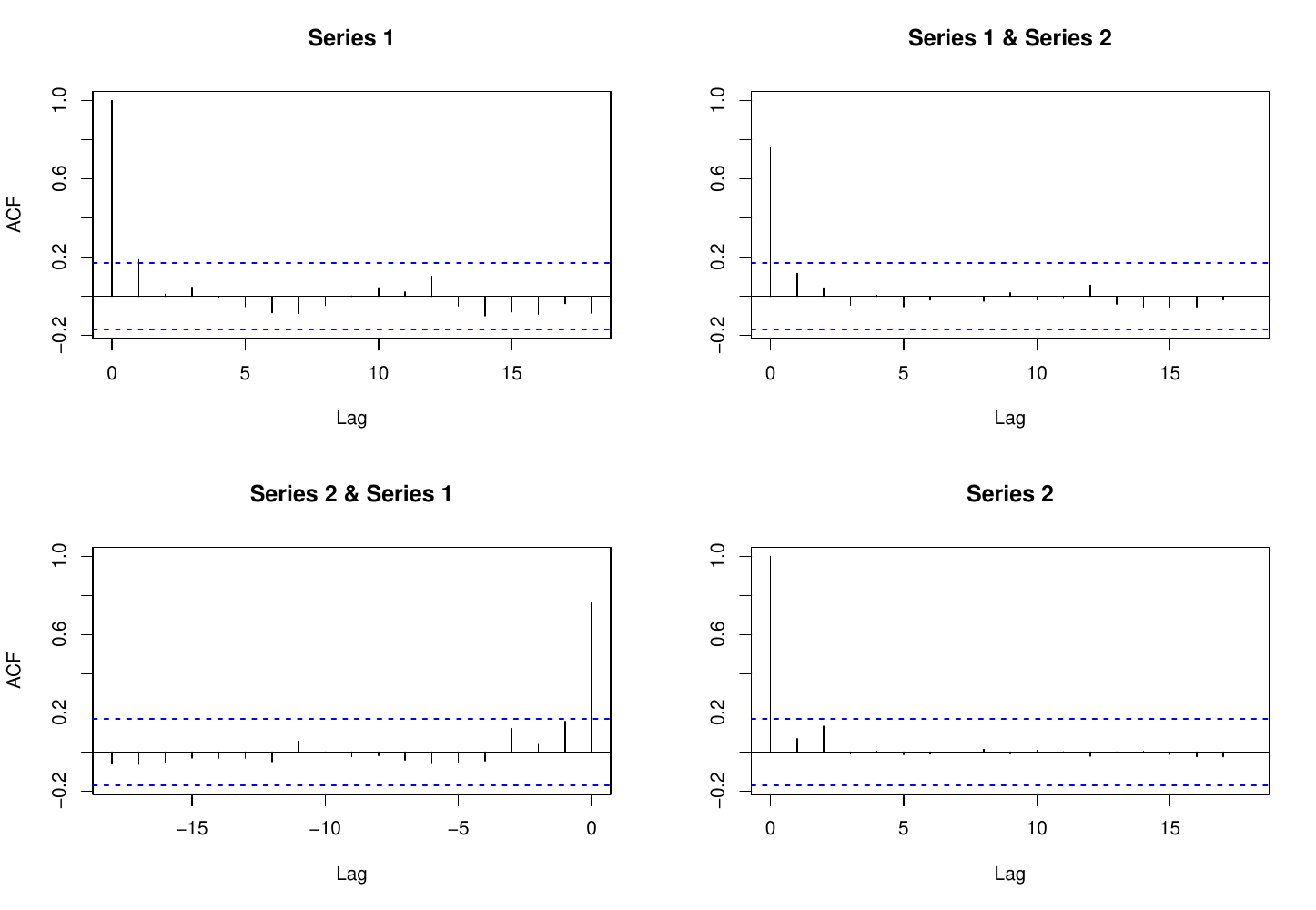}
    \caption{ACF of $\hat{\varepsilon}^2$}
  \end{subfigure}
  \begin{subfigure}[b]{0.4\linewidth}
  \includegraphics[width=\linewidth]{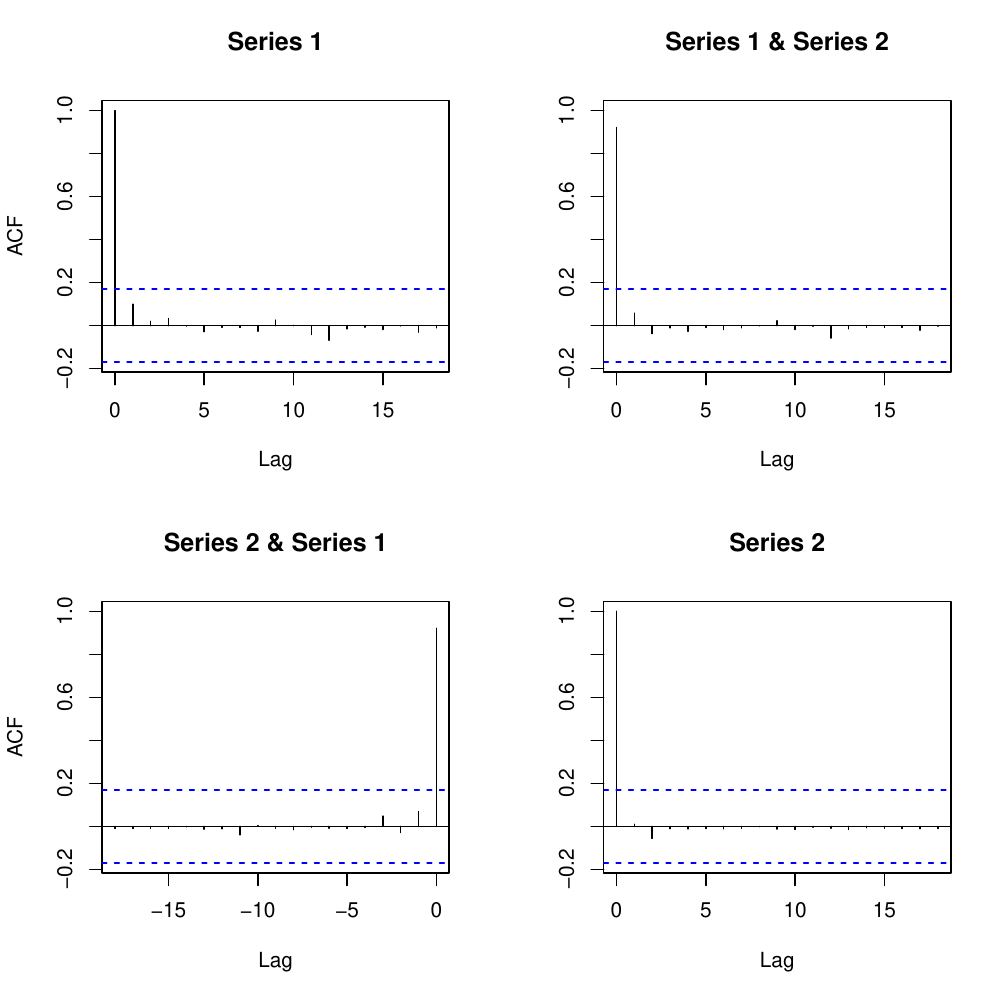}
   \caption{ACF of $\hat{\varepsilon}^3$}
  \end{subfigure}
\caption{Residual ACF}
\end{figure}

We find that the estimated residuals, their squares and third powers  are serially uncorrelated (Figure a.6). The error variance-covariance matrix is 
$\hat{\Sigma} = \left[ \begin{array}{cc}  0.262 & 0.184 \\ 0.184 & 1.506 \end{array} \right]$
and contemporaneous correlation is  0.28 (statistically significant at 0.05).

\nin {\bf 2. Predictive Densities}

We provide the plots of predictive densities of the estimated mixed VAR(1) model of oil prices and US GDP rate computed oos, one step-ahead and conditional on selected dates between T=112 and T=118 marked by the shaded area in Figures a.7-a.11.

\clearpage
\begin{figure}[h]
\centering
\includegraphics[width=12cm,height =12cm, angle = 0]{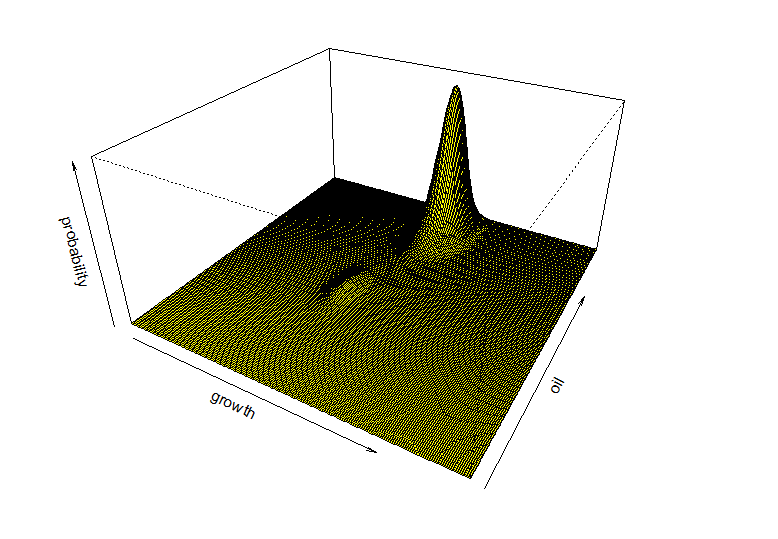}
\caption{VAR(1) based predictive density estimation, T=112}
\end{figure}

\clearpage
\begin{figure}[h]
\centering
\includegraphics[width=15cm,height =12cm, angle = 0]{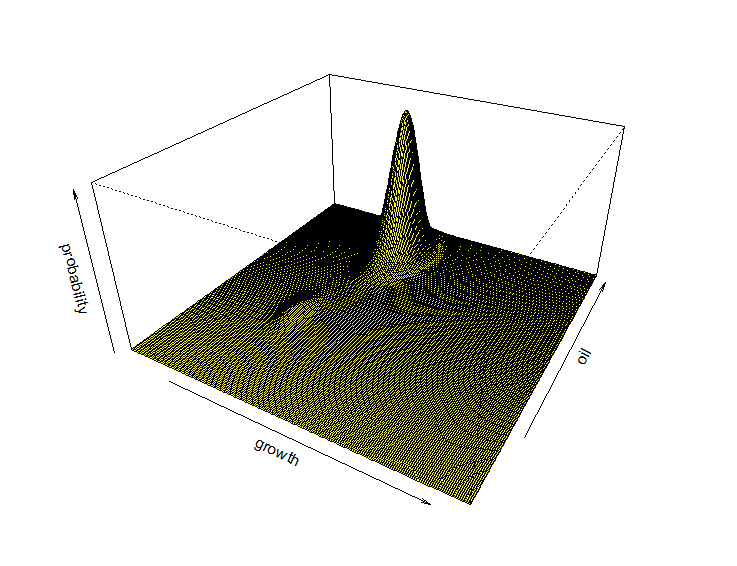}
\caption{VAR(1) based predictive density estimation, T=114}
\end{figure}

\clearpage
\begin{figure}[h]
\centering
\includegraphics[width=15cm,height =12cm, angle = 0]{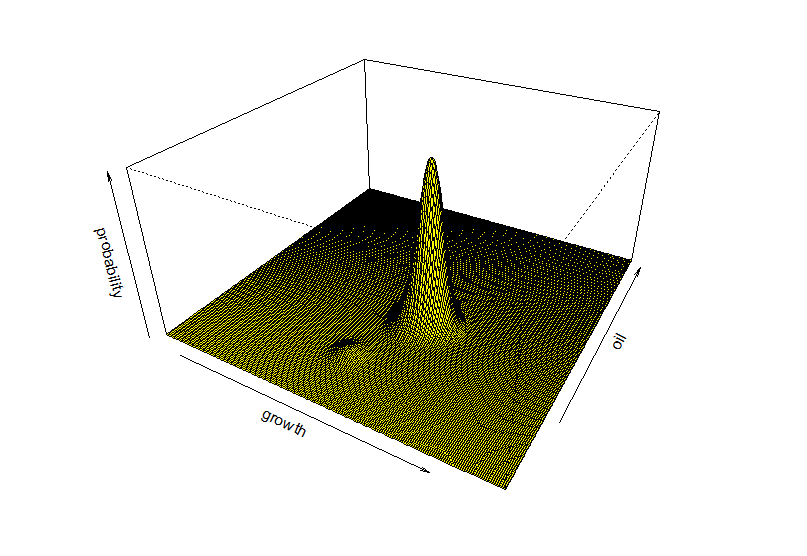}
\caption{VAR(1) based predictive density estimation, T=115}
\end{figure}

\clearpage
\begin{figure}[h]
\centering
\includegraphics[width=12cm,height =12cm, angle = 0]{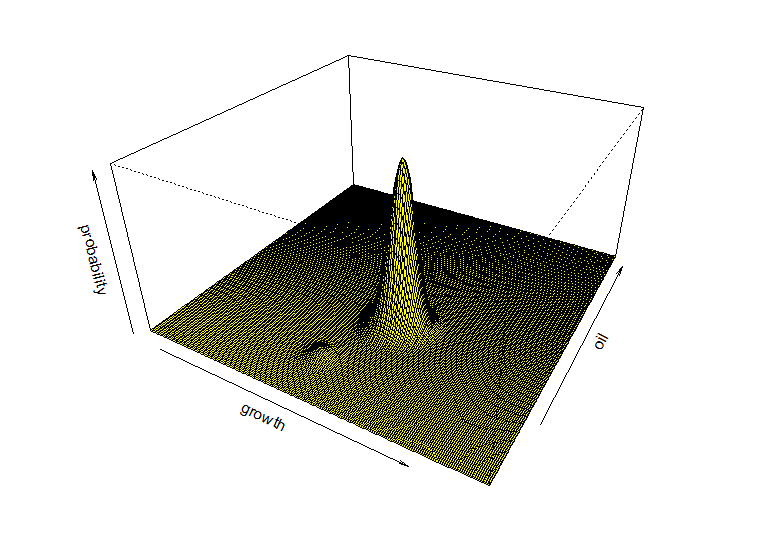}
\caption{VAR(1) based predictive density estimation, T=116}
\end{figure}

\clearpage
\begin{figure}[h]
\centering
\includegraphics[width=12cm,height =12cm, angle = 0]{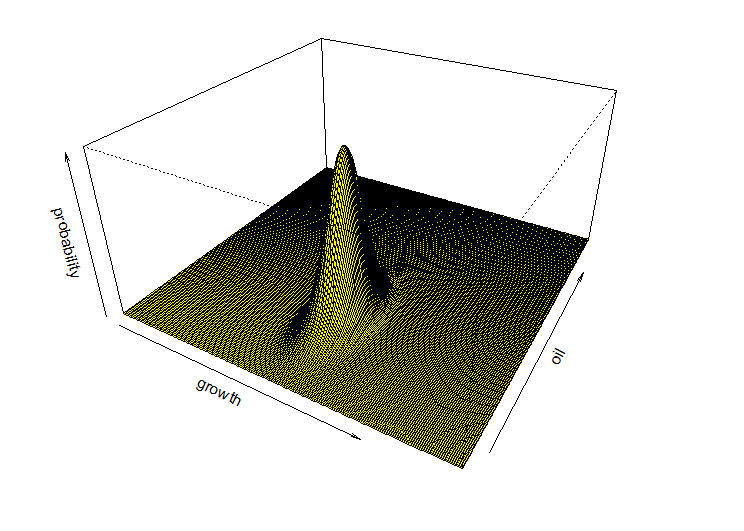}
\caption{VAR(1) based predictive density estimation, T=118}
\end{figure}

We observe deformations during the bubble episode arising due to increasing downturn probabilities, which is consistent with the patterns of univariate predictive densities documented in Hencic et al. (2020), Hecq et al. (2022) and the theoretical results of Freis, Zakoian (2019).

\end{document}